\documentclass[12pt]{article}
\usepackage{epsfig,amssymb,amsmath,psfrag}

\def \bl  {\begin{align*}}
\def \el  {\end{align*}}

\def \be  {\begin{equation}}
\def \ee  {\end{equation}}
\def \ba  {\begin{eqnarray}}
\def \ea  {\end{eqnarray}}
\def \baa {\begin{eqnarray*}}
\def \eaa {\end{eqnarray*}}
\def \bb  {\begin {thebibliography} }
\def \eb  {\end{thebibliography}}
\def \lab #1 {\label{#1}}

\def \qqquad {\qquad\quad}

\def \matrix #1 {\left(\begin{array}{cc} #1 \end{array}\right)}

\renewcommand{\a}{\alpha}
\newcommand{\adt}{{\dot{\alpha}}}
\newcommand{\bdt}{{\dot{\beta}}}
\renewcommand{\b}{\beta}
\newcommand{\lam}{\lambda}
\newcommand{\tlam}{\tilde{\lambda}}
\renewcommand{\AA}{\mathcal{A}}
\newcommand{\BB}{\mathcal{B}}
\newcommand{\CC}{\mathcal{C}}
\newcommand{\cZ}{\mathcal{Z}}
\newcommand{\cW}{\mathcal{W}}

\newcommand \widebar [1] {\overline{#1}}

\def\XXint#1#2#3{{\setbox0=\hbox{$#1{#2#3}{\int}$}
     \vcenter{\hbox{$#2#3$}}\kern-.5\wd0}}

\def\l<{\langle}\def\r>{\rangle}

\begin{document}

\thispagestyle{empty}
\null\vskip-12pt \hfill  LAPTH-029/10 \\
\vskip2.2truecm
\begin{center}
\vskip 0.2truecm {\Large\bf
{\Large Hidden Simplicity of Gauge Theory Amplitudes}
}\\
\vskip 1truecm
{\bf J.~M. Drummond$^{*}$  \\
}

\vskip 0.4truecm
$^{*}$ {\it
LAPTH\footnote{Laboratoire d'Annecy-le-Vieux de Physique Th\'{e}orique, UMR 5108}, Universit\'{e} de Savoie, CNRS\\
B.P. 110,  F-74941 Annecy-le-Vieux Cedex, France\\
\vskip .2truecm                        }
\end{center}

\vskip 1truecm 
\centerline{\bf Abstract} 
These notes were given as lectures at the CERN Winter School on Supergravity, Strings and Gauge Theory 2010. We describe the structure of scattering amplitudes in gauge theories, focussing on the maximally supersymmetric theory to highlight the hidden symmetries which appear. Using the BCFW recursion relations we solve for the tree-level S-matrix in $\mathcal{N}=4$ super Yang-Mills theory, and describe how it produces a sum of invariants of a large symmetry algebra. We review amplitudes in the planar theory beyond tree-level, describing the connection between amplitudes and Wilson loops, and discuss the implications of the hidden symmetries.

\medskip

 \noindent

\newpage
\setcounter{page}{1}\setcounter{footnote}{0}


\section{Introduction}

There are many reasons for studying scattering amplitudes in gauge theories. An obvious current motivation is the need to understand QCD processes in sufficient detail to distinguish new physics from the otherwise overwhelming background at the Large Hadron Collider. The study of the S-matrix can help in the search for new tools for realising this program. Many of the techniques for efficient calculation of scattering amplitudes were first developed using the maximally supersymmetric Yang-Mills theory as a testing ground (see e.g. \cite{Bern:1994zx}). The amplitudes of this theory are similar in general structure to QCD amplitudes but simpler. Moreover they are sufficiently non-trivial to reveal many interesting and surprising mathematical structures governing the behaviour of the S-matrix. 

In fact the planar $\mathcal{N}=4$ theory exhibits an infinite-dimensional symmetry, known as Yangian symmetry which is generated by two distinct versions of superconformal symmetry. The existence of an infinite-dimensional symmetry is a reflection of the integrable structure which is believed to govern physical quantities of the planar theory. This leads to the hope that a solution of the planar S-matrix might be found in this theory which would be a remarkable example of solvability in an interacting four-dimensional gauge theory. This in itself provides another reason for studying the S-matrix of gauge theories.

These lectures will focus on the second motivation. The aim is to review some of the developments that have occurred in recent years with the focus on the symmetries of the S-matrix of $\mathcal{N}=4$ super Yang-Mills theory. However some of the techniques that we will discuss in these notes have wider application than the $\mathcal{N}=4$ theory, in particular the use of recursion relations to derive tree-level amplitudes. Moreover it is to be hoped that the observed symmetries underlying planar $\mathcal{N}=4$ super Yang-Mills theory will lead to a better understanding of the S-matrix in gauge theory more generally. One of the main lessons learned from the existence of recursive techniques and extended symmetries is that manifest locality obscures the general underlying structure of the S-matrix. This is a statement which is not tied to the supersymmetry of the underlying gauge theory.

We will present the solution for the tree-level S-matrix obtained from recursion relations and discuss the appearance of non-trivial symmetries. We will then go on to review the structure of the planar perturbative expansion. Much progress has also been made on the structure of the S-matrix at strong coupling which is accessible in this theory via the AdS/CFT conjecture. There is much overlap between the two regimes of the theory, in particular in the way extended symmetries make an appearance, constraining the form of the S-matrix, and in a relation between the amplitudes and certain light-like Wilson loops. We refer the reader to \cite{Alday:2007hr,Alday:2009yn,Alday:2009dv,Alday:2010vh} for more details of the progress made in solving for the amplitudes at strong coupling.

We will begin by reviewing a few basic properties of tree-level scattering amplitudes in gauge theories in general in Section \ref{trees}. We will discuss the colour structure, helicity structure and general analytic properties of such amplitudes. More details on these topics can be found in the lecture notes of Dixon \cite{Dixon:1996wi}.
We will also introduce $\mathcal{N}=4$ on-shell supersymmetry as we will focus on this theory for most of the course. We will then go on to show in Section \ref{BCFW} how amplitudes can be solved for recursively by exploiting their analytic structure. As we will see, in the $\mathcal{N}=4$ theory this will lead to a complete solution for the tree-level S-matrix. The explicit form of the S-matrix of $\mathcal{N}=4$ super Yang-Mills theory will reveal an unexpected symmetry, namely dual superconformal symmetry which we discuss in detail in Section \ref{dualsconf}. We will see that the full symmetry is the Yangian of the superconformal algebra. Finally Section \ref{loops} will be a brief review of amplitudes at loop level. In particular we discuss the relation to Wilson loops and the way the extended symmetry is exhibited beyond tree level.

\section{Tree-level gauge theory scattering amplitudes}
\label{trees}

We will begin by considering pure Yang Mills theory. The action of this theory is 
\be
S =  -\int d^4 x {\rm Tr}\Bigl(\tfrac{1}{4} F^{\mu \nu}(x) F_{\mu \nu}(x)\Bigr),
\label{YMaction}
\ee
where 
\be
F_{\mu\nu} = \partial_\mu A_\nu - \partial_\nu A_{\mu} - ig^2 [A_\mu,A_\nu].
\ee
Here $A_\mu = A_\mu^a T^a$ and $F_{\mu \nu}^a T^a$ are the Yang-Mills connection and curvature respectively, while $T^a$ are the generators of the gauge group.
The action of course has a gauge symmetry - this is the price of making the locality of the theory manifest. 
In order to write down Feynman rules the gauge should be fixed. In the end we will be interested in gauge-invariant on-shell amplitudes so the choice of gauge will not matter, however it does mean that individual Feynman diagrams are not gauge invariant. In order to compute amplitudes  using Feynman diagrams we must therefore carry around a lot of intrinsically unphysical information which in the end will cancel out. The consequence is that intermediate expressions for even the tree-level amplitudes are much more complicated than the final results.

All of this suggests that there is a better way of expressing scattering amplitudes which does not refer directly to the existence of the gauge-invariant action (\ref{YMaction}). We will indeed see that once we drop the requirement that amplitudes are built from rules derived from a local action we will gain a huge simplicity in the form of the scattering amplitudes. Moreover we will find that new symmetries are revealed which are simply invisible at the level of the Yang-Mills action. 

\begin{figure}
\psfrag{prop}[cc][cc]{$\,\,\,\,\,\,\frac{1}{p^2}$}
\psfrag{3pt}[cc][cc]{$\,\,\,\,\,\,\,f^{abc}\, p^\mu$}
\psfrag{4pt}[cc][cc]{$\,\,\,\,\,\,\,\,\,f^{abc}f^{cde}$}
 \centerline{{\epsfysize5.5cm
\epsfbox{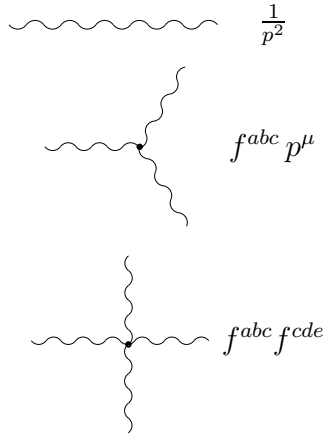}}}  \caption[]{\small
Schematic form for the propagators and vertices for Yang-Mills theory. The detailed index structure is not important for us. We just note that the colour structure enters only via the structure constants, while the momentum dependence appears in the propagators as $1/p^2$ and in the three-point vertices as a positive power of $p$.}
  \label{propsandvertices}
\end{figure}

Let us begin by thinking about the analytic structure of the tree-level diagrams generated by the Yang-Mills action. For simplicity we will choose to work in Feynman gauge $\partial_\mu A^\mu = 0$. In this gauge the propagators and vertices are particularly simple. We show them in Fig. \ref{propsandvertices}.

In order to compute a leading-order on-shell gluon scattering amplitude we then consider all amputated tree-level diagrams contributing to the correlation function,
\be
\langle A^{a_1}_{\mu_1}(p_1) \ldots A^{a_n}_{\mu_n}(p_n) \rangle.
\label{Acorrelator}
\ee

There are various aspects of the calculation which need to be organised. Firstly let us deal with the colour structure. We will be interested in $SU(N)$ gauge groups. Every instance of the $SU(N)$ structure constants $f^{abc}$ can be written in terms of the generators
\be
f^{abc} = {\rm Tr}(T^a T^b T^c) - {\rm Tr}(T^a T^c T^b).
\ee

\begin{figure}
\psfrag{rep}[cc][cc]{$\longrightarrow$}
\psfrag{-}[cc][cc]{$-$}
\psfrag{+}[cc][cc]{$-\frac{1}{N}$}
 \centerline{{\epsfysize5.5cm
\epsfbox{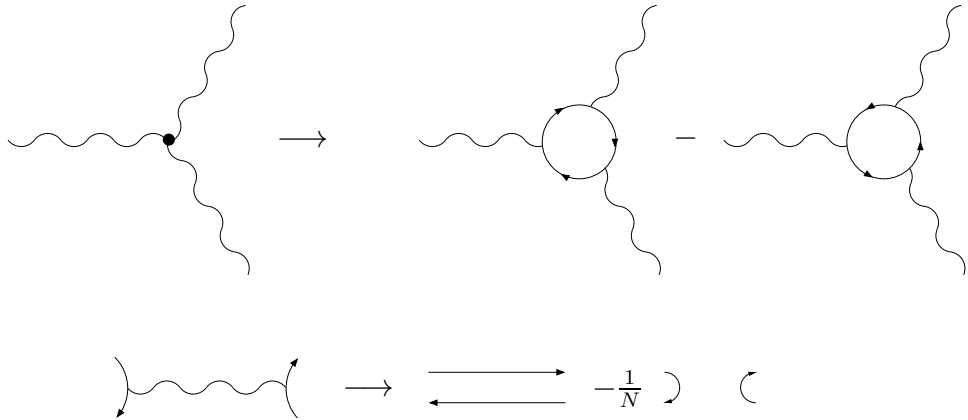}}}  \caption[]{\small
  Colour rules.}
  \label{colour}
\end{figure}

When two adjoint indices $a$ and $b$ are contracted via a propagator we can use the relation
\be
T^{a}_{i}{}^{\bar i} T^{a}_{j}{}^{\bar j} = \delta_i^{\bar j} \delta_j^{\bar i} - \frac{1}{N} \delta_i^{\bar{i}} \delta_j^{\bar{j}}.
\label{suNgens}
\ee
These replacements can be represented diagrammatically as in Fig. \ref{colour}.

Now if we take into account all propagators and vertices and vertices in a diagram we find that we end up with one term which is a single trace going round the whole tree-level diagram plus other terms which are related to this one by swapping over all external legs in all possible ways (the $1/N$ terms from (\ref{suNgens}) always drop out). Thus in the end the amplitude can be expressed as a sum over non-cyclic permutations of a single cyclically ordered partial amplitude $\mathcal{A}_{\mu_1,\ldots,\mu_n}(p_1,\ldots,p_n)$,
\be
\langle A_{\mu_1}^{a_1} \ldots A_{\mu_n}^{a_n} \rangle = \sum_{\sigma \in S_n / Z_n} {\rm Tr} (T^{a_{\sigma(1)}} \ldots T^{a_{\sigma(n)}}) \mathcal{A}_{\mu_1 \ldots \mu_n}(p_{\sigma(1)},\ldots,p_{\sigma(n)}).
\label{colourdecomp}
\ee

Let us consider the physical on-shell degrees of freedom of Yang-Mills theory. This is most easily captured by the spinor helicity formalism. A four-momentum $p_i^\mu$ can be thought of as a bispinor after contraction with spin matrices $(\sigma_\mu)^{\alpha \dot\alpha}$,
\be
p_i^{\alpha \dot\alpha}  = (\sigma_\mu)^{\alpha \dot \alpha} p_i^\mu.
\ee
The square $p_i^2$ is then the determinant of the $2 \times 2$ matrix $p^{\alpha \dot\alpha}$.
The on-shell condition then says that the momentum can be expressed as the product of two commuting spinors
\be
\det (p^{\alpha \dot\alpha})=0 \iff p_i^{\alpha \dot\alpha}  = \lambda_i^{\alpha} \tilde{\lambda}_i^{\dot \alpha}.
\label{spinordef}
\ee
Here $\tlam^\adt = \pm \bar{\lam}^\adt = \pm (\lam^\a)^*$ with the sign given by the sign of the energy component of $p$. Note that the $\lam$ and $\tlam$ are not unambiguously fixed by the definition (\ref{spinordef}). The ambiguity is given by the freedom to rescale by a phase,
\be
\lam \longrightarrow e^{i \phi} \lam, \qquad \tlam \longrightarrow e^{-i \phi} \tlam.
\ee
This rescaling is generated by the helicity operator,
\be
h = \frac{1}{2}\Bigl[-\lam^\a \frac{\partial}{\partial \lam^\a} + \tlam^\adt \frac{\partial}{\partial \tlam^\adt}\Bigr].
\ee
By convention we have chosen $\lam$ to have helicity $-\tfrac{1}{2}$ and $\tlam$ to have helicity $\tfrac{1}{2}$.

The free Yang-Mills field equations $\partial^\mu F_{\mu\nu}=0$ and Bianchi identity $\partial_{[\mu}F_{\nu \rho]}=0$ are expressed in the two-component notation as 
\be
\partial^{\alpha \dot\alpha} F_{\alpha \beta} = 0, \qquad \partial^{\alpha \dot\alpha} F_{\dot\alpha \dot\beta} =0,
\label{eomandbianchi}
\ee
where $F_{\alpha \beta}$ and $F_{\dot\alpha \dot\beta}$ are the self-dual and anti-self-dual parts of the field strength respectively,
\be
F_{\alpha \dot{\alpha} \beta \dot{\beta}} = F_{\alpha \beta}\epsilon_{\dot{\alpha} \dot{\beta}} + F_{\dot{\alpha} \dot{\beta}} \epsilon_{\alpha \beta}.
\ee
Note that $F_{\alpha \beta}$ and $F_{\dot{\alpha}\dot{\beta}}$ are symmetric in their indices. 

The equations (\ref{eomandbianchi}) can be written in terms of the momentum $p^{\alpha \dot\alpha}=\lambda^\alpha \tilde{\lambda}^{\dot\alpha}$ as
\begin{align}
\lambda^\alpha \tilde{\lambda}^{\dot\alpha} F_{\a \b} &= 0 \implies F_{\a \b} = \lam_\a \lam_\b G_+, \\
\lam^\a \tlam^{\adt} F_{\adt \bdt} &= 0 \implies F_{\adt \bdt} = \tlam_{\adt} \tlam_\bdt G_-.
\end{align}
Thus we see that the Yang-Mills equations admit two on-shell solutions described by $G_+$ and $G_-$, the positive and negative helicity gluon states, carrying helicity $1$ and $-1$ respectively (so that $F$ has no helicity). 

We are interested in the scattering of these on-shell states so our amplitudes will be characterised by an ordered sequence of $+$ and $-$ signs and we will have an on-shell momentum for each particle,
\be
p_i^{\alpha \dot\alpha} = \lam_i^\a \tlam_i^\adt.
\ee
Our conventions will be that all particles are incoming. Of course an incoming positive helicity particle is equivalent to an outgoing negative helicity particle and vice-versa.

The colour-ordered, partial amplitudes can be obtained from $\mathcal{A}_{\mu_1 ... \mu_n}$ of (\ref{colourdecomp}) by contracting each of the Lorentz indices with the appropriate polarisation vector. The polarisation vectors can be defined with the help of auxiliary light-like momenta $l_i^{\a \adt} = \mu_i^\a \tilde{\mu_i}^\adt$,
\be
\epsilon_{+,i}^{\a \adt} = \frac{\tlam_i^\adt \mu_i^\a}{\l< \lam_i \mu_i \r>}, \qquad \epsilon_{-,i}^{\a \adt} = \frac{\lam_i^\a \tilde{\mu_i}^\adt}{[\tlam_i \tilde{\mu_i}]}.
\ee
Here we have introduced the notation for the spinor scalar products,
\be
\l< a b \r> = a^\a b_\a = a^\a b^\b \epsilon_{\b \a}, \qquad [a b] = a_\adt b^\adt = a_\adt b_\bdt \epsilon^{\adt \bdt}.
\ee
So an ordered amplitude is given by, for example
\be
\mathcal{A}(+,+,-,\ldots,+,-) = \epsilon_{+,1}^{\mu_1} \epsilon_{+,2}^{\mu_2} \epsilon_{-,3}^{\mu_3} \ldots\epsilon_{+,n-1}^{\mu_{n-1}} \epsilon_{-,n}^{\mu_n} \mathcal{A}_{\mu_1 \ldots \mu_n}(p_1,\ldots,p_n).
\ee
The amplitude does not depend on the auxiliary momenta $\mu_i$ used to define the polarisation vector. A shift in $\mu_i$ is simply a gauge transformation. For example, under the shift $\mu \longrightarrow \mu + \delta \mu$ we have
\be
\delta \epsilon_+^{\alpha \dot\alpha} = \frac{\tlam^\adt \delta \mu^\a}{\l< \lam \mu \r>} - \tlam^\adt \mu^\a \frac{\l<\lam \delta \mu \r>}{\l< \lam \mu \r>^2} = \frac{\tlam^\adt \delta \mu^\a \l<\lam \mu \r> - \tlam^{\adt} \mu^\a \l< \lam \delta \mu \r>}{\l< \lam \mu \r>^2}.
\ee
Using the cyclic identity $a^\a \l< bc \r> + b^\a \l< ca \r> + c^\a \l< ab \r> = 0$ we have
\be
\delta \epsilon_+^{\a \adt} = \lam^\a \tlam^\adt \frac{\l< \delta \mu \mu \r>}{\l< \lam \mu \r>^2} = p^{\a \adt} \frac{\l< \delta \mu \mu \r>}{\l< \lam \mu \r>^2}.
\ee
The overall factor of $p^{\a \adt}$ means that the variation of the polarisation vector contributes nothing to the amplitude due to the Ward identity
\be
p^\mu \l< A_\mu(p)  \ldots \r> = 0.
\ee
Thus the amplitudes depend on the variables $\{ \lam_i, \tlam_i\}$ only.

When expressed in momentum space, amplitudes will always have an overall factor of $\delta^4(p) = \delta^4(p_1+\ldots +p_n)$ as a consequence of translation invariance,
\be
\mathcal{A}_n = \delta^4(p) A_n.
\ee
After stripping off the overall momentum conserving delta function, we can think of the scattering amplitudes as being given by a single Lorentz-invariant rational function $A_n$ of the ordered set of spinors $\{\lam_1,\tlam_1 \ldots, \lam_n, \tlam_n\}$ with only {\sl local} poles of the form,
\begin{align}
\frac{1}{(p_i + p_{i+1} + \ldots + p_j)^2}. 
\end{align}
The poles originate from the propagators in the Feynman diagram expansion. The fact that the momenta in the denominator form an ordered sum $(p_i + p_{i+1} + \ldots + p_j)$ is due to the fact that we are considering the ordered partial amplitude. The presence and structure of the poles are crucial analytic properties which we will need in order to be able to solve for all tree-level amplitudes. 

\begin{figure}
\psfrag{MHV}[cc][cc]{MHV}
\psfrag{NMHV}[cc][cc]{NMHV}
\psfrag{NNMHV}[cc][cc]{NNMHV}
\psfrag{MHVb}[cc][cc]{$\overline{\rm MHV}$}
\psfrag{4pt}[cc][cc]{$n=4$}
\psfrag{5pt}[cc][cc]{$n=5$}
\psfrag{6pt}[cc][cc]{$n=6$}
\psfrag{Parity}[cc][cc]{Parity}
 \centerline{{\epsfysize8cm
\epsfbox{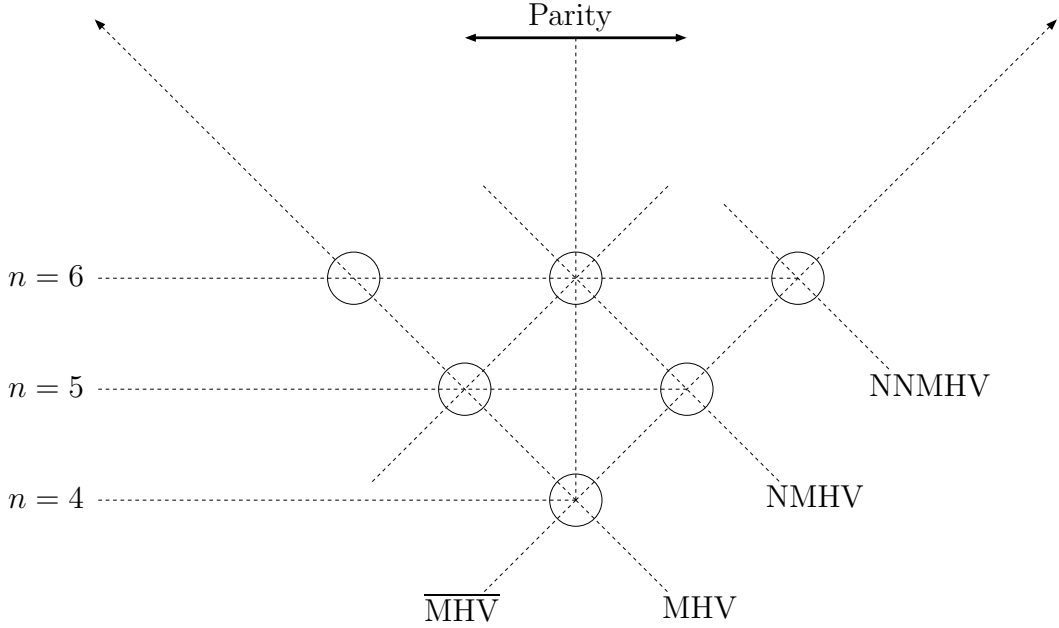}}}  \caption[]{\small
  Different amplitudes classified according to their helicity structures. Parity acts as reflection about the vertical axis of the diagram, swapping MHV and $\overline{\rm MHV}$ amplitudes for example.}
  \label{MHVexpansion}
\end{figure}

We can classify amplitudes according to their helicity structure. At tree level the amplitudes with no negative helicity gluon or only one negative helicity gluon vanish. This can be seen by making a suitable choice of polarisation vectors but we will shortly see a symmetry-based argument for why this is the case.
Those with two negative helicity gluons and the rest positive helicity are called the maximally helicity-violating (MHV) amplitudes. Those with three negative helicities are called next-to-MHV (NMHV) and so on. The names come from the fact that a helicity-conserving amplitude would have the same incoming and outgoing helicity structure. In the convention with all incoming particles this means an equal number of positive and negative helicity particles (obviously this would require an even number of particles in total). Therefore the helicity configuration furthest away from an equal number is called maximally-helicity-violating. 
By parity the amplitudes must also have at least two positive helicity gluons (those with exactly two are called the anti-MHV or $\overline{\rm MHV}$ amplitudes). The simplest non-trivial amplitude is therefore the four-particle amplitude which is both MHV and $\overline{\rm MHV}$.
The classification is illustrated in Fig. \ref{MHVexpansion}.

So far we have discussed pure Yang-Mills theory where the only scattering particles are the two gluon states. More general gauge theories will have additional particles describing the particular matter content of the theory. At tree level the pure gluon scattering amplitudes however are common to every gauge theory, regardless of the matter content. This is simply because the only Feynman diagrams which arise in their calculation are the ones with gluons on every line. There is one gauge theory which surpasses all others in its remarkable properties which is the maximally supersymmetric one, $\mathcal{N}=4$ super Yang-Mills theory. The theory has sixteen on-shell states, eight bosons and eight fermions. The bosonic states are the two helicity states of the gluon and six real scalars which transform in the adjoint representation of $su(4)$ (or vector of $so(6)$). The fermionic states are the four gluinos and four anti-gluinos transforming in the anti-fundamental and fundamental representation of $su(4)$ respectively,
\be
\text{bosons: } \quad G_+, \quad G_-, \quad S_{AB} = \tfrac{1}{2} \epsilon_{ABCD} S^{CD}, \qquad \text{fermions: } \quad \Gamma_A, \quad \overline{\Gamma}^A. 
\ee

The $\mathcal{N}=4$ theory is unique in that all on-shell states arrange themselves into a single PCT self-conjugate multiplet. We can describe this multiplet by a superfield depending on four Grassmann variables $\eta^A$, transforming in the fundamental representation of $su(4)$,
\be
\Phi(\eta) = G_+ + \eta^A \Gamma_A + \tfrac{1}{2!} \eta^A\eta^B S_{AB} + \tfrac{1}{3!} \eta^A \eta^B \eta^C \epsilon_{ABCD} \overline{\Gamma}^D + \tfrac{1}{4!} (\eta)^4 G_-.
\label{etarep}
\ee
Moreover, if we assign helicity $\tfrac{1}{2}$ to the variable $\eta$ we see that the whole superfield $\Phi$ has helicity 1. In other words we have added a term to the helicity generator,
\be
h = \frac{1}{2}\Bigl[-\lam^\a \frac{\partial}{\partial \lam^\a} + \tlam^\adt \frac{\partial}{\partial \tlam^\adt} + \eta^A \frac{\partial}{\partial \eta^A}\Bigr], 
\ee
so that
\be
h \Phi(\eta) = \Phi(\eta).
\ee
We have made a choice in writing (\ref{etarep}) by putting the positive helicity gluon in the bottom component of the multiplet. We could equally well have written the multiplet the other way round by expanding in a Grassmann variable in the anti-fundamental representation of $su(4)$,
\be
\bar{\Phi}(\bar\eta) = G_- + \bar{\eta}_A \overline{\Gamma}^A + \tfrac{1}{2!} \bar{\eta}_A\bar{\eta}_B S^{AB} + \tfrac{1}{3!} \bar{\eta}_A \bar{\eta}_B \bar{\eta}_C \epsilon^{ABCD} \Gamma_D + \tfrac{1}{4!} (\bar{\eta})^4 G_+.
\ee
The two superfields are related by a Grassmann Fourier transform,
\be
\bar{\Phi}(\bar\eta) = \int d^4 \eta\, e^{\eta \cdot \bar\eta} \Phi(\eta).
\label{GFT}
\ee

The supersymmetry generators take the form
\be
p^{\a \adt} = \lam^\a \tlam^{\adt}, \qquad q^{\a A} = \lam^\a \eta^A, \qquad \bar{q}^{\adt}_A = \tlam^{\adt} \frac{\partial}{\partial \eta^A}.
\label{pqbarq}
\ee
It is straightforward to see that these generators, together with the Lorentz and $su(4)$ generators,
\be
M_{\a \b} = \lam_{(\a} \frac{\partial}{\partial \lam^{\b)}}, \qquad \overline{M}_{\adt \bdt} = \tlam_{(\adt} \frac{\partial}{\partial \tlam^{\bdt)}}, \qquad R^{A}{}_{B} = \eta^A \frac{\partial}{\partial \eta^B} - \frac{1}{4} \delta^A_B \eta^C \frac{\partial}{\partial \eta^C},
\label{mbarmr}
\ee
do indeed give a representation of the super Poincar\'e algebra. 

The fact that the $\mathcal{N}=4$ theory is PCT self-conjugate means that all $n$-particle amplitudes arrange themselves into a single superamplitude. All of the component amplitudes can then be obtained by expanding in the Grassmann variables,
\be
\mathcal{A}(\Phi_1,\ldots,\Phi_n) = (\eta_1)^4 (\eta_2)^4 \mathcal{A}(+,+,-,-,\ldots,-) + (\eta_1)^4 (\eta_2)^3 \eta_3 \mathcal{A}(+,\overline{\Gamma}, \Gamma, -,\ldots,-)  + \ldots.
\ee
We have suppressed the $su(4)$ indices in the second term for simplicity. The amplitudes have helicity 1 in each particle,
\be
h_i \mathcal{A}(\Phi_1, \ldots , \Phi_n) = \mathcal{A}(\Phi_1, \ldots ,\Phi_n).
\ee
When we consider the superamplitudes the symmetry generators are simply the sums of the single-particle representations (\ref{pqbarq},\ref{mbarmr}). For example we have
\be
p^{\a \adt} = \sum_i \lam_i^\a \tlam_i^\adt, \qquad q^{\a A} = \sum_i \lam_i^\a \eta_i^A, \qquad \bar{q}^{\adt}_A = \sum_i \tlam_i^{\adt} \frac{\partial}{\partial \eta_i^A}.
\label{superPoincaremulti}
\ee

What can the symmetries of the theory tell us about the scattering amplitudes? Assuming that the momentum variables $\lam$ and $\tlam$ are not constrained, translation invariance tells us there is an overall momentum conserving delta function $\delta^4(p)$. Similarly supersymmetry tells us that there should be a delta function $\delta^8(q)$. We recall that the delta function of a Grassmann quantity $\delta(\psi)$ is simply $\psi$ itself. Thus we have that the amplitude can be written,
\be
\mathcal{A}(\Phi_1,\ldots,\Phi_n) = \frac{\delta^4(p) \delta^8(q)}{\langle 12 \rangle  \ldots \langle n1 \rangle} \mathcal{P}_n(\lam,\tlam,\eta).
\label{superamplitude}
\ee
Here we have put the factor $\langle 12 \rangle \langle 23 \rangle \ldots \langle n-1 \, n \rangle \langle n1 \rangle$ in the denominator to carry the helicities of the superparticles so that the function $\mathcal{P}_n$ has helicity zero for every particle, i.e. it is annihilated by each $h_i$.

The fact that there is a factor $\delta^8(q)$ in (\ref{superamplitude}) due to supersymmetry means that that certain component amplitudes (e.g. pure gluon amplitudes with fewer than two negative helicity gluons) must vanish because there at least eight $\eta$ variables in (\ref{superamplitude}). This statement about the vanishing of certain pure gluon scattering amplitudes is true in all gauge theories at tree level because such amplitudes are common to every gauge theory regardless of the matter content. This shows that all gauge theories exhibit the effects of $\mathcal{N}=4$ supersymmetry, even if they are not supersymmetric theories. Of course the statement is true in $\mathcal{N}=4$ super Yang-Mills even beyond tree-level since it is a consequence of supersymmetry while in other gauge theories the all-plus amplitudes, for example, are non-vanishing at loop level.

The function $\mathcal{P}_n$ is also constrained by symmetry. In particular the $su(4)$ symmetry means that Grassmann variables always appear in multiples of four so that $\mathcal{P}_n(\lam,\tlam,\eta)$ can be expanded into terms of Grassmann degree 0,4,8 etc. These terms correspond to the classification of amplitudes as MHV, NMHV, NNMHV and so on,
\be
\mathcal{P}_n(\lam,\tlam,\eta) = \mathcal{P}_n^{(0)} + \mathcal{P}_n^{(4)} + \mathcal{P}_n^{(8)} + \ldots + \mathcal{P}_n^{(4n-16)}.
\ee
The final term in the expansion of $\mathcal{P}_n$ corresponds to the $\overline{\rm MHV}$ amplitudes. The $\bar{q}$ supersymmetry also imposes constraints on the form of $\mathcal{P}_n$. Indeed we have $\bar{q}^{\adt}_A \mathcal{P}_n=0$. Note that we can use the $\bar{q}$ supersymmetry to fix any two of the $\eta$ variables to zero. The finite $\bar{q}$ transformation with parameter given by
\be
\xi^A_\adt = \frac{\tlam_{i\adt} \eta_j^A - \tlam_{j\adt} \eta_i^A}{[\tlam_1 \tlam_n]}
\ee
will set $\eta_i$ and $\eta_j$ to zero. We will use this fact when we consider recursive relations among tree-level amplitudes in the $\mathcal{N}=4$ theory.

The symmetries (\ref{pqbarq},\ref{mbarmr}) are not the only symmetries of the the theory. $\mathcal{N}=4$ super Yang-Mills theory is a superconformal field theory and the dilatation generator,
\be
d = \frac{1}{2}\sum_i\Bigl[ \lam_i^\a \frac{\partial}{\partial \lam_i^\a} + \tlam_i^{\adt} \frac{\partial}{\partial \tlam_i^{\adt}} \Bigr],
\ee
and the special conformal and superconformal generators, 
\be
k_{\a \adt} = \sum_i \frac{\partial^2}{\partial \lambda_i^\a \partial \tlam_i^\adt}, \qquad s_{\a A} = \sum_i \frac{\partial^2}{\partial \lam_i^\a \partial \eta_i^A}, \qquad \bar{s}_{\adt}^A = \sum_i \eta_i^A \frac{\partial}{\partial \tlam_i^{\adt}}
\label{sconf}
\ee
are also symmetries of the tree-level amplitudes \cite{Witten:2003nn}. We will see that even this large symmetry algebra is not the end of the story. When written in the most compact way the tree-level amplitudes reveal another superconformal symmetry, called dual superconformal symmetry.

\section{BCFW recursion relations}
\label{BCFW}

We will now see how we can reconstruct the entire tree-level S-matrix from the simple analytic structure that we have described in the previous section. We will first present the general recursive method which is due to Britto, Cachzo, Feng and Witten \cite{Britto:2004ap,Britto:2005fq}. The presentation of the method will essentially follow that of \cite{Britto:2005fq}. The arguments can be framed in a very general form, in particular they can be applied to theories in any number of dimensions \cite{ArkaniHamed:2008yf}. Here we will focus on the case of four dimensions and make direct use of the spinor helicity formalism. We will then formulate the method in a supersymmetric fashion as in \cite{Brandhuber:2008pf,ArkaniHamed:2008gz} and then use it to solve for the tree-level S-matrix of $\mathcal{N}=4$ super Yang-Mills \cite{Drummond:2008cr}.

We will consider a tree-level gluon amplitude with incoming massless momenta $p_i^{\a \adt} = \lam_i^\a \tlam_i^\adt$. We will consider deforming the momenta by making the following shift of the spinor variables,
\begin{align}
\lam_1 \longrightarrow \hat{\lam}_1(z) &= \lam_1 - z \lam_n, \notag \\
\tlam_n \longrightarrow \hat{\tlam}_n(z) &= \tlam_n + z \tlam_1,
\end{align}
where $z$ is a complex number\footnote{We are choosing to shift $\lam_1$ and $\tlam_n$ for later convenience but one can derive relations for amplitudes by shifting any pair of legs.}. Under this shift the momenta $p_1$ and $p_n$ are deformed in a complex direction by an amount proportional to $z$,
\begin{align}
p_1^{\a \adt} \longrightarrow \hat{p}_1^{\a \adt}(z) &= (\lam_1^\a - z \lam_n^\a) \tlam_1^{\adt},\notag\\
p_n^{\a \adt} \longrightarrow \hat{p}_n^{\a \adt}(z) &= \lam_n^{\a}(\tlam_n^{\adt} + z \tlam_1^{\adt}).
\end{align}
Note that by construction the shifted momenta are still light-like $\hat{p}_1^2 = \hat{p}_n^2=0$ and that momentum is still conserved, $\hat{p}_1 + \hat{p_n} = p_1 + p_n$.

\begin{figure}
\psfrag{hat1}[cc][cc]{$\hat{1}$}
\psfrag{hatn}[cc][cc]{$\hat{n}$}
\psfrag{Shifted}[cc][cc]{shifted propagators}
\psfrag{MHVb}[cc][cc]{$\overline{\rm MHV}$}
 \centerline{{\epsfysize5cm
\epsfbox{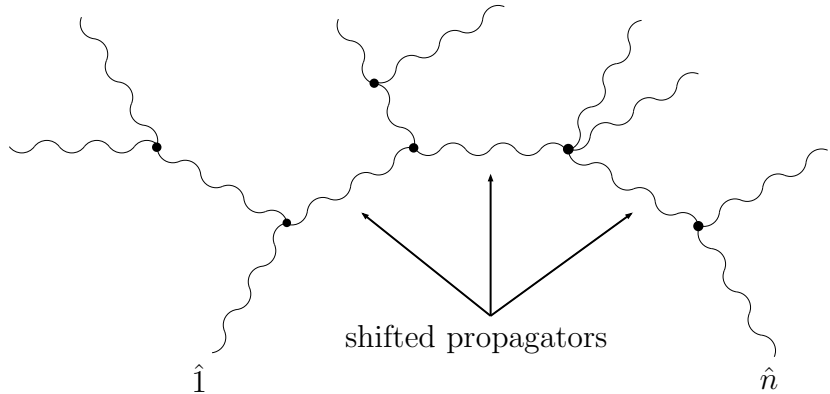}}}  \caption[]{\small
  An example of a Feynman diagram showing the propagators affected by the BCFW shift.}
  \label{shiftedprops}
\end{figure}

As we have deformed the momenta $p_1$ and $p_n$ the amplitude will also be deformed to become a function of $z$. What can we say about the analytic structure of the amplitude as a function of $z$? We have already seen that we can write a given amplitude as 
\be
\mathcal{A}_n = \delta^4(p) A_n,
\ee
where $A_n$ is a rational function of the spinor variables $\{\lam_i, \tlam_i\}$ with only local poles of the form
\be
\frac{1}{(p_i + p_{i+1} + \ldots p_j)^2}.
\label{localpole}
\ee
This implies that the deformed amplitude $A(z)$ will only have simple poles as a function of $z$. The only propagator factors which will exhibit a dependence on $z$ are those of the form
\be
\frac{1}{(p_1 + p_2 + \ldots + p_{i-1})^2} \equiv \frac{1}{P_i^2}.
\label{ithpole}
\ee
For simplicity of notation we will simply write $P$ instead of $P_i$ until we need to remember that there are many such poles. The propagators which are affected by the shift in one example are shown in Fig. \ref{shiftedprops}.

\begin{figure}
\psfrag{hat1}[cc][cc]{$\hat{1}$}
\psfrag{hatn}[cc][cc]{$\hat{n}$}
\psfrag{AL}[cc][cc]{\,\,$\mathcal{A}_L$}
\psfrag{AR}[cc][cc]{\,\,\,$\mathcal{A}_R$}
\psfrag{i-1}[cc][cc]{$i-1$}
\psfrag{i}[cc][cc]{$i$}
\psfrag{s}[cc][cc]{$s$}
\psfrag{sb}[cc][cc]{$\bar{s}$}
\psfrag{sum}[cc][cc]{$\sum_{s}$}
 \centerline{{\epsfysize5cm
\epsfbox{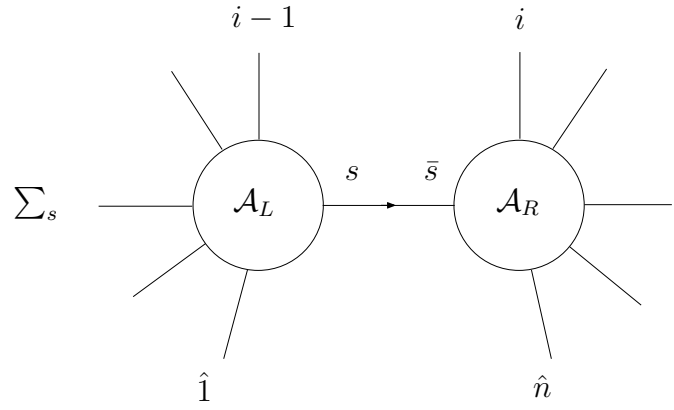}}}  \caption[]{\small
  The sum over states giving a particular residue.}
  \label{statesum}
\end{figure}

Under the $z$-shift such a pole will become
\be
\frac{1}{P^2} \longrightarrow \frac{1}{\hat{P}(z)^2} = \frac{1}{(\hat{p}_1(z) +p_2 + \ldots p_{i-1})^2} = \frac{1}{P^2 - z \langle n | P | 1]}.
\ee
Here we adopt the notation that $\l< n | P | 1] = \lambda_n^\a P_{\a \adt} \tlam_1^{\adt}$. We have found that the amplitude has a pole at 
\be
z_{P} = \frac{P^2}{\langle n | P |1]}
\ee
for every possible propagator of the form (\ref{ithpole}).
At this value of $z$ the complex shift that we have performed is such that the internal propagator carrying momentum $\hat{P}(z)$ has gone on shell,
\be
\hat{P}(z_{P})^2 = 0.
\ee
Near the pole the amplitude behaves as follows
\be
A_n(z) \sim \frac{1}{(z-z_P)} \biggl(\frac{-1}{\langle n|P |1]} \biggr) \sum_s A_L^s \bigl( \hat{1}(z_P),\ldots,i-1,-\hat{P}(z_P) \bigr) A_R^{\bar{s}} \bigl( \hat{P}(z_P),i,\ldots,\hat{n}(z_P) \bigr).
\ee
This notation means that when the intermediate propagator goes on-shell, every diagram factorises into left and right pieces, with every external leg on shell. Summing up all diagrams which possess the internal propagator in question one obtains tree-level subamplitudes $A_L^s$ and $A_R^{\bar{s}}$ on either side of the intermediate on-shell propagator. Since one adds all possible Feynman diagrams, the state $s$ exchanged between the two subamplitudes can be anything and one must sum over all possible states. In pure Yang-Mills theory this means that either a positive or a negative helicity gluon can be exchanged over the internal line. In $\mathcal{N}=4$ super Yang-Mills theory then the exchanged state can be any of the sixteen on-shell states of the theory. The sum over subamplitudes contributing to a particular residue is illustrated in Fig. \ref{statesum}.

\begin{figure}
\psfrag{=}[cc][cc]{$=$}
\psfrag{+res}[cc][cc]{$\qquad \qquad + \,\, {\rm res}(z=\infty)$}
 \centerline{{\epsfysize4.5cm
\epsfbox{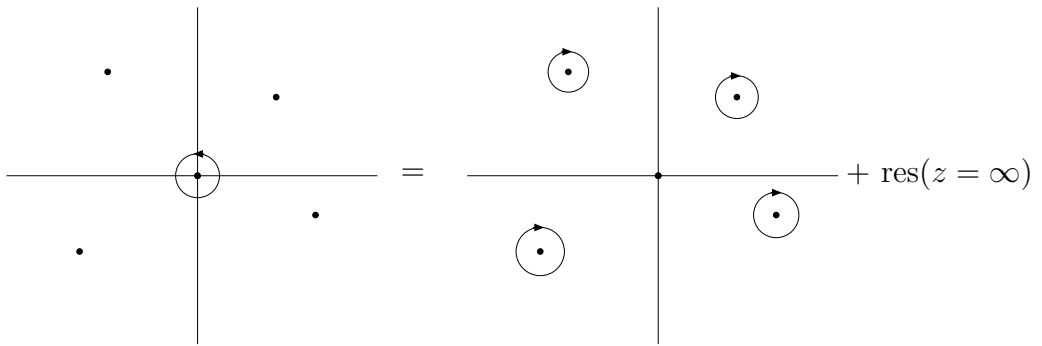}}}  \caption[]{\small
  The contour deformation giving a sum over residues.}
  \label{contour}
\end{figure}

Let us now consider the function $A(z)/z$. Near the pole at $z=z_P$ we find
\be
\frac{A(z)}{z} \sim - \frac{1}{z-z_P} A_L(z_P) \biggl( \frac{1}{P^2} \biggr) A_R(z_P).
\ee
The initial amplitude we are interested in can be written as
\be
A = A(0) = \oint_C \frac{dz}{2 \pi i} \frac{A(z)}{z}.
\ee
Here we have chosen the contour to be a small circle around the origin. By deforming the contour we can write the amplitude as a sum over residues from the other poles plus a potential contribution from infinity. This contour deformation is pictured in Fig. \ref{contour}.

Thus we obtain
\be
A(0) = \sum_{i} \sum_s A_L^s(z_{P_i}) \biggl(\frac{1}{P_i^2}\biggr) A_R^{\bar{s}}(z_{P_i}) \, + \, \text{res($z=\infty$)},
\ee
where we have restored the notation $P_i$ to refer to the different possible poles of the form (\ref{ithpole}) which can arise.
For the amplitudes we are interested in we will find that the contribution from $z=\infty$ vanishes and so we have the BCFW recursion relation
\be
A(0) = \sum_{i} \sum_s A_L^s(z_{P_i}) \biggl(\frac{1}{P_i^2}\biggr) A_R^{\bar{s}}(z_{P_i}).
\ee

\begin{figure}
\psfrag{hat1}[cc][cc]{$\hat{1}$}
\psfrag{hatn}[cc][cc]{$\hat{n}$}
\psfrag{z}[cc][cc]{$z$}
\psfrag{1/z}[cc][cc]{$\frac{1}{z}$}
 \centerline{{\epsfysize5cm
\epsfbox{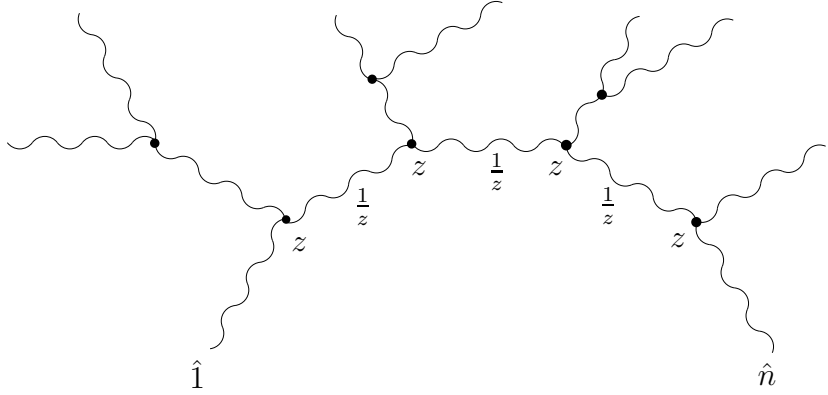}}}  \caption[]{\small
  An example of a Feynman diagram contributing to the leading $z$ dependence as $z$ goes to infinity. The vertices joining the line of shifted propagators are all three-point vertices.}
  \label{zdependence}
\end{figure}

Let us now consider how the function $A(z)$ behaves as $z$ goes to infinity. To address this question we again need to refer back to the Feynman rules. Let us consider how a typical Feynman diagram behaves under the shift we are performing. Between the legs carrying the shifted momenta $\hat{p}_1$ and $\hat{p}_n$ there will be a succession of internal propagators joining vertices which connect to the unshifted part of the diagram. Each internal propagator which feels the shift will contribute a negative power of $z$ as $z\longrightarrow \infty$. Any three-point vertices on the line of the shifted momenta contribute a positive power of $z$ (due to the fact that the three-point gluon coupling contains a derivative of the gluon field). Any four point vertices are milder, contributing no $z$ dependence. Therefore the dominant behaviour as $z \longrightarrow \infty$ comes from diagrams where the vertices along the line of shifted propagators are all three-point interactions, as in Fig. \ref{zdependence}. There is always one more vertex than internal propagator so we conclude that the dominant Feynman diagrams scale like $z$ as $z\longrightarrow \infty$. 

We must also include the effects of the polarisation vectors. Since they are momentum dependent, the polarisation vectors for particles $1$ and $n$ can also contribute to the large $z$ behaviour. The scalings of the possible choices for positive and negative helicity gluons are summarised below,
\begin{align}
\epsilon_{1,+}^{\a \adt} &= \frac{\tilde{\lambda}_1^{\dot\alpha} \mu^\alpha}{\l< \hat{\lam}_1(z) \mu \r>} \sim \frac{1}{z}, \qquad \epsilon_{n,+}^{\a \adt} = \frac{\hat{\tlam}_n^{\adt}(z) \mu^\a}{\l<\lam_n \mu\r>}  \sim  z, \notag \\
\epsilon_{1,-}^{\a \adt} &= \frac{ \hat{\lam}_1^{\a}(z) \tilde{\mu}^{\adt}}{[\tlam_1 \tilde{\mu}]} \,\sim \, z, \qquad \epsilon_{n,-}^{\a \adt} = \frac{\lam_n^\a \tilde{\mu}^{\adt}}{[\hat{\tlam}_n(z) \tilde{\mu}]} \sim \frac{1}{z}.
\label{polarisation1n}
\end{align}
Thus we see that the dominant contributions to each amplitude scale differently depending on the choice of the helicities of the shifted legs. In summary we have the following limits on the large $z$ behaviour (noting only the helicities for particles 1 and $n$),
\be
\mathcal{A}(+-) \sim \frac{1}{z}, \qquad \mathcal{A}(++) \sim z, \qquad \mathcal{A}(--) \sim z, \qquad \mathcal{A}(-+) \sim z^3.
\label{largezlimits}
\ee
The fact that the $(+-)$ amplitude falls off as $z\longrightarrow \infty$ means that it is always possible to choose legs corresponding to this helicity configuration in order to get a recursive relation to lower-point amplitudes. In fact we will need a stronger result than this in order to proceed with finding the solution for tree-level scattering amplitudes in the $\mathcal{N}=4$ theory. We would like to show that the large $z$ limiting behaviour (\ref{largezlimits}) is improved for the case of the $(++)$ and $(--)$ amplitudes.
Note that in (\ref{largezlimits}) we only have limits on the asymptotic behaviour because it could happen that in certain cases there is a cancellation among the contributions coming from different diagrams resulting in much softer large $z$ behaviour. Indeed we will now argue that in the case of the $(++)$ and $(--)$ amplitudes this is exactly what happens, leading to a suppression by two extra powers of $z$. The argument closely follows the discussion of Arkani-Hamed and Kaplan \cite{ArkaniHamed:2008yf}. 

In the limit of large $z$ we can think of the scattering amplitude as the amplitude for a single particle at very large (complex) momentum scattering of some soft background describing the other particles. So let us consider the Lagrangian expanded around some soft background. We will write the gauge field as $A_{\mu} = B_\mu + a_\mu$ where $B_\mu$ is the soft background field and $a_\mu$ is the fluctuation. Adding the gauge-fixing term $(D_\mu a^\mu)^2$ to the Lagrangian we find the terms quadratic in $a_\mu$ are given by
\be
\mathcal{L}^{\rm quad} = -\frac{1}{4} {\rm Tr} D_\mu a_\nu D^\mu a^\nu + \frac{i}{2} {\rm Tr} G^{\mu \nu} [a_\mu,a_\nu].
\ee
Here $G_{\mu\nu}$ is the field strength for the background field $B_\mu$. The first term contains the derivative couplings and is responsible for the leading behaviour at large $z$. It also has a symmetry which is broken only by the second term in the quadratic Lagrangian. The symmetry is a Lorentz symmetry which acts only on the indices of the fluctuation field $a_\mu$ but not on the indices of the background fields or derivatives. This symmetry is referred to as `spin-Lorentz' symmetry in \cite{ArkaniHamed:2008yf}. To make it explicit we will use Latin characters for the relevant indices,
\be
\mathcal{L}^{\rm quad} = -\frac{1}{4} {\rm Tr} D_\mu a_a D^\mu a_b \eta^{ab} + \frac{i}{2}{\rm Tr} G^{ab} [a_a, a_b].
\ee
Here we see that the leading term is invariant under Lorentz transformations of the Latin indices while the second term breaks this symmetry due to the presence of the antisymmetric tensor $G^{ab}$. Thus the leading contribution to the two-point function will be proportional to $\eta^{ab}$ while the next correction will be given when there is exactly one coupling to the background field $G^{ab}$ and hence will be antisymmetric in the spin-Lorentz indices $a$ and $b$. Corrections given by two or more couplings to the background will have a generic tensor structure in the spin-Lorentz indices. In summary the two-point function for the hard particle in the soft background will behave as follows,
\be
\mathcal{A}^{ab} = \eta^{ab}(c z + \ldots) + A^{ab} + \frac{1}{z} B^{ab} + \ldots , 
\ee
where $A^{ab}$ is an antisymmetric tensor in the spin-Lorentz indices. In two-component notation we can write this as follows,
\be
\mathcal{A}^{\a \adt \b \bdt} = \epsilon^{\a \b} \epsilon^{\adt \bdt} (c z + \ldots) + (\epsilon^{\a \b} \tilde{s}^{\adt \bdt} + \epsilon^{\adt \bdt} s^{\a \b}) + \frac{1}{z} B^{\a \adt \b \bdt} + \ldots,
\label{background2ptfn}
\ee
where $s^{\a \b}$ and $\tilde{s}^{\adt \bdt}$ are both symmetric in their indices. If we now contract this expression with the polarisation vectors $\epsilon_{1,+}^{\a \adt}$ and $\epsilon_{n,+}^{\b \bdt}$ from (\ref{polarisation1n}) we can see that the leading term in (\ref{background2ptfn}) does not contribute. The subleading term in (\ref{background2ptfn}) contributes the following term to $\mathcal{A}(++)$,
\be
\frac{[\tlam_1 \hat{\tlam}_n(z)] s^{\a \b} \mu_\a \mu_\b}{\l< \hat{\lam}_1(z) \mu \r> \l< \lam_n \mu\r>}.
\ee
The $z$ dependence in the numerator actually drops out because $[\tlam_1 \hat{\tlam}_n(z)] = [\tlam_1 \tlam_n]$. Thus we see that in fact the $(++)$ amplitude is suppressed by two powers of $z$ relative to the worst Feynman diagrams. The same happens for the $(--)$ amplitude. Recall that an incoming positive-helicity particle can be thought of as an outgoing negative-helicity one and vice versa. So if one thinks of one of the particles as incoming and the other as outgoing then the $(++)$ amplitudes and $(--)$ amplitudes we are discussing actually correspond to a single particle which flips its helicity when scattering off the soft background. Physically we can therefore think of the extra suppression by two powers of $z$ as a penalty for the hard particle for flipping its helicity while scattering off the soft background.

In summary we have found the following improved behaviour for the scattering amplitudes at large $z$,
\be
\mathcal{A}(+-) \sim \frac{1}{z}, \qquad \mathcal{A}(++) \sim \frac{1}{z}, \qquad \mathcal{A}(--) \sim \frac{1}{z}, \qquad \mathcal{A}(-+) \sim z^3.
\ee

The fact that the $(++)$ amplitude falls off as $z$ goes to infinity is crucial. 
Recall that in the $\mathcal{N}=4$ theory we can use a $\bar{q}$-supersymmetry transformation to shift the $\eta$ variables for any two legs to zero. If we use this transformation to shift the $\eta$ variables associated to the shifted legs we can relate the full superamplitude to the $(++)$ amplitude. In order to make use of $\bar{q}$-supersymmetry we should perform the shift in the $\mathcal{N}=4$ theory in a way which is compatible $\bar{q}$ transformations. Recall that $\bar{q}$-supersymmetry relates the $\eta$ and $\tlam$ variables so if we are to respect $\bar{q}$-supersymmetry then we should shift $\eta$ whenever we shift $\tlam$. Thus the full shift we will perform in the $\mathcal{N}=4$ theory is the following \cite{Brandhuber:2008pf,ArkaniHamed:2008gz},
\begin{align}
\lam_1 \longrightarrow \hat{\lam}_1(z) &= \lam_1 - z \lam_n, \notag \\
\tlam_n \longrightarrow \hat{\tlam}_n(z) &= \tlam_n + z \tlam_1, \notag\\
\eta_n \longrightarrow \hat{\eta}_n(z) &= \eta_n + z \eta_1.
\end{align}
With this definition of the shift we can see that the parameter of the finite $\bar{q}$ transformation required to set $\eta_1$ and $\hat{\eta}_n(z)$ to zero is independent of $z$,
\be
\xi^A_{\adt} = \frac{\tlam_{1\adt} \hat{\eta}_n^A(z) - \hat{\tlam}_{n \adt}(z) \eta_1^A}{[\tlam_1 \hat{\tlam}_n(z)]} = \frac{\tlam_{1\adt} \eta_n^A - \tlam_{n \adt} \eta_1^A}{[\tlam_1 \tlam_n]}.
\ee
Thus by using $\bar{q}$-supersymmetry we can relate the $z$-dependence of the whole superamplitude to that of its $(--)$ component (where the helicities relate to particles $1$ and $n$ as before). Thus we see that with the correct supersymmetric definition of the shift the whole superamplitude falls off like $1/z$ as $z$ goes to infinity \cite{ArkaniHamed:2008gz}.

So for the superamplitudes in the $\mathcal{N}=4$ theory we have a recursion relation with no contribution from $z=\infty$. Furthermore the sum over states can be replaced by a single Grassmann integral over the $\eta$ variable associated to the internal line joining the two subamplitudes in the recursion relation. In summary the recursion relation for $\mathcal{N}=4$ super Yang-Mills theory is
\be
A_n = \sum_i \int \frac{d^4 \eta_{\hat{P}_i}}{P_i^2} A_L\bigl( \hat{1}(z_{P_i}),2,\ldots,i-1,-\hat{P}(z_{P_i})\bigr) A_R \bigl( \hat{P}(z_{P_i}),i,\ldots,n-1,\hat{n}(z_{P_i}) \bigr).
\label{superBCFW}
\ee

We will proceed to solving the recursion relation to obtain the full tree-level S-matrix for $\mathcal{N}=4$ super Yang-Mills theory. To start the recursion we will need on-shell three-point vertices. The reason that these are needed is that it can happen that the internal propagator closest to leg $1$ (or leg $n$) is actually attached to a three point vertex with both external legs 1 and 2 (or $n-1$ and $n$). The contribution from the pole when this type of internal propagator goes on shell will involve on-shell three-point vertices with complex incoming momenta. The fact that the momenta are complex is important because for real momenta the three-point amplitude vanishes. Let us consider on-shell three-point kinematics. Momentum conservation reads
\be
\lam_1^\a \tlam_1^\adt + \lam_2^\a \tlam_2^\adt = - \lam_3 \tlam_3^\adt .
\label{3ptmomcons}
\ee
Taking the square of both sides tells us
\be
\l< 12 \r> [12] = 0
\ee
and hence
\be
\l< 12 \r> = 0 \qquad \text{ or } \qquad [12]=0.
\ee
If $\l< 12 \r> =0$ then $\lam_1 \propto \lam_2$ and then (\ref{3ptmomcons}) tells us that $\lam_1 \propto \lam_2 \propto \lam_3$ and hence all of the possible angle brackets vanish. Likewise if $[12]=0$ all the square brackets vanish so we have two distinct possibilities for three-point vertices, which we will call $\overline{\rm MHV}$ and MHV,
\begin{align}
&\l< 12 \r> = \l< 23 \r> = \l<31 \r> = 0 \qquad (\overline{\rm MHV}), \\
&[12] \,=\, [23] \, = \, [31] \,=0 \qquad \text{(MHV)}.
\end{align}
Note that for real momenta both conditions are satisfied because the $\lam_i$ and $\tlam_i$ are related by complex conjugation thus the particle momenta would all have to be collinear. This is why the amplitude vanishes for real momenta.

For complex momenta we can construct the three-point amplitudes from their symmetries.
Let us consider the MHV case first. We need to find a three-point superamplitude with helicity 1 on each leg. Translation invariance and $q$-supersymmetry tell us that there are factors of $\delta^4(p)$ and $\delta^8(q)$ as before while Lorentz invariance and the helicity conditions uniquely fix the other factors so that we obtain
\be
\mathcal{A}_3^{\rm MHV} = \frac{\delta^4(p) \delta^8(q)}{\l< 12 \r> \l<23 \r> \l< 31 \r>}.
\ee
The $\overline{\rm MHV}$ case is related to this one by parity. To find the amplitude we interchange $\lam_i$ and $\tlam_i$ and replace $\eta_i$ by $\bar{\eta}_i$. Then to express the amplitude back in terms of the $\eta_i$ we perform the Grassmann Fourier transform (\ref{GFT}) for each leg. The result is \cite{Brandhuber:2008pf,ArkaniHamed:2008gz}
\be
\mathcal{A}_3^{\overline{\rm MHV}} = \frac{\delta^4(p) \delta^4(\eta_1 [23] + \eta_2 [31] + \eta_3 [12])}{[12][23][31]}.
\label{3ptMHVbar}
\ee
It may be slightly surprising that for this amplitude $q$-supersymmetry does not imply that there is a factor of $\delta^8(q)$ as usual. The reason is that the three-point kinematics in the $\overline{\rm MHV}$ case are such that all $\lam$ variables are parallel. This means that $q^{\a A}$ itself factorises, $q^{\a A} = \lam_F^\a q_F^A$ for some $\lam_F^\a$ and $q_F^A$ and the requirement of $q$-supersymmetry is only that the amplitude contain a factor of $\delta^4(q_F)$ as we indeed find in (\ref{3ptMHVbar}). Thus the $\overline{\rm MHV}_3$ amplitude has Grassmann degree four while all other amplitudes have Grassmann degree at least eight.

The recursion relation (\ref{superBCFW}) can be decomposed into contributions of various Grassmann degrees. Since there is a Grassmann integral on the RHS the sum of the degrees of the two subamplitudes $A_L$ and $A_R$ must be four more than the degree of the amplitude we are solving for. The us we find
\begin{eqnarray}
\label{super-BCF-all} 
A_{n}^{{\rm
N}^{p}{\rm MHV}} &=& 
\int \frac{d^{4}
\eta_{\hat{P}}}{P^2} \, A_{3}^{\rm \widebar{MHV}}(z_{P})
A_{n-1}^{{\rm N}^{p}{\rm MHV}}(z_{P}) 
\nonumber \\
 &+& 
 \sum_{m=0}^{p-1} \; \; \sum_i  
\int \frac{d^{4}\eta_{\hat{P}_{i}}}{P_i^2}
A_{i}^{{\rm N}^{m}{\rm MHV}}(z_{P_{i}})
A_{n-i+2}^{{\rm N}^{(p-m-1)}{\rm MHV}}(z_{P_{i}})\,.
\end{eqnarray}
Note that we have not allowed for the left subamplitude to be $A^{\rm MHV}_3$. This cannot happen because in the MHV case the square bracket $[12]$ vanishes. For the left subamplitude the $\tlam$ variables are unshifted and hence this would imply that $[12]$ and hence $(p_1+p_2)^2$ vanishes for the full amplitude as well. This is a restriction on the kinematics which is not true in general and hence such a term does not contribute to the recursion relation. Similarly the right subamplitude can never be $A^{\overline{\rm MHV}}_3$.

\begin{figure}
\psfrag{hat1}[cc][cc]{$\hat{1}$}
\psfrag{hat4}[cc][cc]{$\hat{4}$}
\psfrag{2}[cc][cc]{$2$}
\psfrag{3}[cc][cc]{$3$}
\psfrag{MHVb}[cc][cc]{$\overline{\rm MHV}$}
\psfrag{MHV}[cc][cc]{MHV}
\psfrag{P}[cc][cc]{$\hat{P}$}
 \centerline{{\epsfysize4cm
\epsfbox{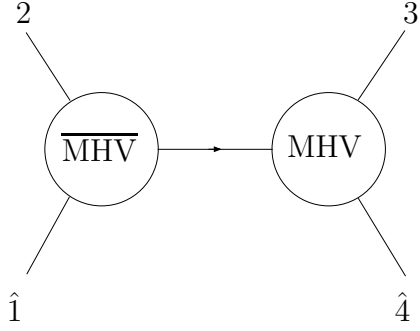}}}  \caption[]{\small The single BCFW diagram contributing to the four-point amplitude.}
  \label{4ptamp}
\end{figure}

We can now begin to construct amplitudes recursively using the three-point amplitudes as the starting point. The simplest amplitude is the four-point amplitude. There is only one pole which arises under the $z$ shift and hence only one term in the recursion relation which is given by
\be
A_4 = \int \frac{d^4\eta_{\hat{P}}}{P^2} A_3^{\overline{\rm MHV}}(\hat{1},2,-\hat{P}) A_3^{\rm MHV}(\hat{P},3,\hat{4}).
\ee
This BCFW term is represented in Fig. \ref{4ptamp}.

Using the form of the three-point amplitudes we find
\be
A_4 = \int \frac{d^4 \eta_{\hat{P}}}{P^2} \frac{\delta^4(\eta_1[2\hat{P}] + \eta_2 [\hat{P} 1] + \eta_{\hat{P}}[12])}{[12][2\hat{P}][\hat{P}1]} \frac{\delta^8(\lam_{\hat{P}} \eta_{\hat{P}} +\lam_3 \eta_3 + \lam_4 \hat{\eta}_4)}{\l< \hat{P} 3 \r> \l< 34 \r> \l< 4\hat{P} \r>}
\ee
The $\delta^4$ factor tells us that $\eta_{\hat{P}}$ can be expressed in terms of the other $\eta$ variables,
\be
\eta_{\hat{P}} = -\frac{1}{[12]} (\eta_1 [2\hat{P}] + \eta_2 [\hat{P} 1]).
\ee
Examining the $\delta^8$ factor we see that it can then be written
\begin{align}
&\delta^8 \bigl( -\lam_{\hat{P}} \frac{1}{[12]}(\eta_1[2\hat{P}] + \eta_2 [\hat{P} 1]) + \lam_3 \eta_3 + \lam_4 \hat{\eta}_4 \bigr)
= \delta^8\bigl( \hat{\lam}_1 \eta_1 + \lam_2 \eta_2 + \lam_3 \eta_3 +\lam_4\hat{\eta}_4 \bigr) = \delta^8(q).
\end{align}
where in the first step we have used momentum conservation to write $\lam_{\hat{P}}\tlam_{\hat{P}} = \hat{\lam}_1 \tlam_1 + \lam_2 \tlam_2$ and in the second we cancel the $z$ dependence between the first and last terms in the argument. Thus we obtain the expected supersymmetry delta function $\delta^8(q)$.

\begin{figure}
\psfrag{hat1}[cc][cc]{$\hat{1}$}
\psfrag{hat4}[cc][cc]{$\hat{n}$}
\psfrag{2}[cc][cc]{$2$}
\psfrag{3}[cc][cc]{$3$}
\psfrag{MHVb}[cc][cc]{$\overline{\rm MHV}$}
\psfrag{MHV}[cc][cc]{MHV}
\psfrag{P}[cc][cc]{$\hat{P}$}
 \centerline{{\epsfysize4cm
\epsfbox{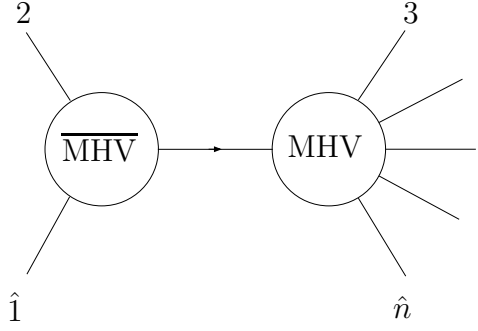}}}  \caption[]{\small The single BCFW diagram contributing to the $n$-point MHV amplitude.}
  \label{nptamp}
\end{figure}

All that remains is to collect the bosonic factors together and simplify them. The $\delta^4$ factor soaks up the Grassmann integration, producing a factor of $[12]^4$ due to the coefficient in front of $\eta_{\hat{P}}$ in the argument. The factor of $P^2$ in the denominator can be written $(p_1 + p_2)^2 = \l< 12 \r> [12]$. Thus we have the bosonic factors
\be
\frac{[12]^4}{\l<12 \r> [12]^2 [2\hat{P}][\hat{P}1] \l<\hat{P} 3\r> \l< 34 \r> \l< 4 \hat{P} \r>} =\frac{[12]^2}{\l< 12 \r> ([\hat{P}1]\l< \hat{P} 3\r>)\l< 34 \r> ([2 \hat{P}]\l< 4 \hat{P} \r>)}
\ee
Again momentum conservation allows us to write
\be
[\hat{P}1]\l< \hat{P} 3\r> = [21]\l< 23 \r>, \qquad [2 \hat{P}]\l< 4 \hat{P} \r> = [21] \l<4 \hat{1}\r> = [21] \l<41 \r>.
\ee
Thus in total we obtain the expected expression for the four-point amplitude
\be
A_4 = \frac{\delta^8(q)}{\l< 12 \r> \l< 23 \r> \l< 34 \r> \l< 41 \r>}.
\ee
An almost identical calculation shows that $n$-point MHV amplitude takes the form
\be
A_n^{\rm MHV} = \frac{\delta^8(q)}{\l< 12 \r> \ldots \l< n1 \r>}.
\label{AMHV}
\ee
Again there is only a single BCFW term, represented in Fig. \ref{nptamp}.
This formula first appeared in \cite{Nair:1988bq}, supersymmetrising the Parke-Taylor formula \cite{Parke:1986gb,Berends:1987me} for MHV gluon amplitudes.

\begin{figure}
\psfrag{dots}[cc][cc]{$\ldots$}
\psfrag{hat1}[cc][cc]{$\hat{1}$}
\psfrag{2}[cc][cc]{$2$}
\psfrag{3}[cc][cc]{$3$}
\psfrag{hatn}[cc][cc]{$\hat{n}$}
\psfrag{i-1}[cc][cc]{$i-1$}
\psfrag{P}[cc][cc]{$\hat{P}$}
\psfrag{Pi}[cc][cc]{$\hat{P}_{i}$}
\psfrag{i}[cc][cc]{\hspace{0cm}$i$}
\psfrag{MHV}[cc][cc]{MHV}
\psfrag{NMHV}[cc][cc]{NMHV}
\psfrag{MHVb}[cc][cc]{$\overline{\rm MHV}$}
\psfrag{+sum}[cc][cc]{$\,\, + \,\, \sum_{i=4}^{n-1}$}
 \centerline{{\epsfysize4cm
\epsfbox{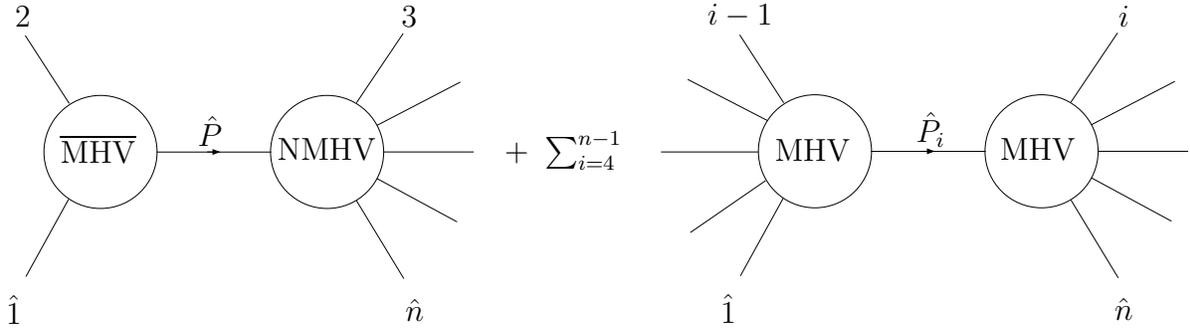}}}  \caption[]{\small
  The two contributions to the supersymmetric recursion relation for
  NMHV amplitudes. We call them homogeneous and
  inhomogeneous respectively.}
  \label{fig1}
\end{figure}

Let us now consider the next-to-MHV (NMHV) case. The recursion relation has two types of terms, one where the subamplitudes are $A^{\overline{\rm MHV}}_3$ and $A_{n-1}^{\rm NMHV}$ and the other where they are both MHV,
\be
A_n^{\rm NMHV} = \int \frac{d^{4}
\eta_{\hat{P}}}{P^2} \, A_{3}^{\rm \widebar{MHV}}(z_{P})
A_{n-1}^{{\rm N}^{p}{\rm MHV}}(z_{P}) + \sum_{i=3}^{n-1} \int \frac{d^{4}\eta_{\hat{P}_{i}}}{P_i^2}
A_{i}^{{\rm MHV}}(z_{P_{i}})
A_{n-i+2}^{{\rm MHV}}(z_{P_{i}})\,.
\label{NMHVBCFW}
\ee
The two kinds of terms are represented in Fig. \ref{fig1}.

Let us look at the second type of term first. We call this the inhomogeneous term because it only involves the MHV amplitudes which we have already solved for. Writing out the $i$th term in the sum in more detail we have
\be
I_i=\int \frac{d^4 \eta_{\hat{P}_{i}}}{P_i^2} \frac{\delta^8\bigl(\hat{\lam}_1 \eta_1 + \sum_2^{i-1} \lam_j \eta_j - \lam_{\hat{P}_i}\eta_{\hat{P}_i}\bigr)}{\l< \hat{1} 2 \r> \l< 23 \r> \ldots \l< i-1 \, \hat{P}_i\r>\l<\hat{P}_i \hat{1}\r>} \frac{\delta^8\bigl(\lam_{\hat{P}_i} \eta_{\hat{P}_i} + \sum_i^{n-1} \lam_j \eta_j - \lam_n \hat{\eta}_n\bigr)}{\l<\hat{P}_i i\r> \l< i \, i+1 \r> \ldots \l<n \hat{P}_i \r>}.
\ee
One of the two $\delta^8$ factors can immediately be exchanged for the overall supersymmetry conserving delta function with the help of the other $\delta^8$,
\be
I_i=\delta^8(q) \int \frac{d^4 \eta_{\hat{P}_i}}{P_i^2} \frac{\delta^8\bigl(\hat{\lam}_1 \eta_1 + \sum_2^{i-1} \lam_j \eta_j - \lam_{\hat{P}_i}\eta_{\hat{P}_i}\bigr)}{\l< \hat{1} 2 \r> \l< 23 \r> \ldots \l< i-1 \, \hat{P}_i\r> \l<\hat{P}_i \hat{1}\r> \l<\hat{P}_i i\r> \l< i+1 \, i+2 \r> \ldots \l<n \hat{P}_i \r>}.
\label{inhomterm}
\ee

We must now perform the Grassmann integral and simplify the remaining bosonic factors. In order to organise the calculation it will be very helpful to introduce `dual' coordinates $x_i$ and $\theta_i$. These are defined so that their differences give the momenta and supercharges associated to the external legs,
\be
x_i^{\a \adt} - x_{i+1}^{\a \adt} = \lam_i^\a \tlam_i^{\adt} = p_i^{\a \adt}, \qquad \theta_i^{\a A} - \theta_{i+1}^{\a A} = \lam_i^\a \eta_i^A = q_i^{\a A}.
\label{dualxandtheta}
\ee
We will give a more geometric interpretation to these dual coordinates in the next section where we will see that they are very helpful in revealing a new symmetry (dual superconformal symmetry). Here we are just using them for notational convenience. We will use the shorthand notation
\be
x_{ij} \equiv x_i - x_j = p_i + p_{i+1} + \ldots + p_{j-1}, \qquad \theta_{ij} \equiv \theta_i - \theta_j = q_i + q_{i+1} + \ldots q_{j-1}.
\ee
Note that as a consequence of their definitions the dual coordinates satisfy some useful relations,
\be
\l<i|x_{ij} = \l<i|x_{i+1,j}, \qquad \l<i|\theta_{ij} = \l<i|\theta_{i+1,j}
\ee
and similarly with angle brackets replaced by square ones.

Returning to (\ref{inhomterm}) we can split the $\delta^8$ into a product of two $\delta^4$ factors,
\be
\delta^8\biggl(\hat{\lam}_1 \eta_1 + \sum_2^{i-1} \lam_j \eta_j - \lam_{\hat{P}_i}\eta_{\hat{P}_i}\biggr) = \l< \hat{1} \hat{P}_i \r>^4 \delta^4\biggl( \eta_1 + \sum_2^{i-1} \frac{\l<\hat{P}_i j \r>}{\l<\hat{P}_i \hat{1} \r>} \eta_j \biggr)\delta^4 \biggl( \sum_2^{i-1} \frac{\l< \hat{1} j \r>}{\l<\hat{1} \hat{P}_i \r>} \eta_j - \eta_{\hat{P}_i} \biggr).
\label{delta4}
\ee
The second $\delta^4$ soaks up the Grassmann integration. The remaining $\delta^4$ and the angle bracket factors combine to give
\be
\delta^4\biggl(\l< \hat{P}_i \hat{1} \r> \eta_1 + \sum_2^{i-1} \l< \hat{P}_i j \r> \eta_j \biggr) = \delta^4\biggl(\sum_1^{i-1} \l< \hat{P}_i j \r> \eta_j - z \l< \hat{P}_i n \r> \eta_1\biggr ).
\ee
Noting that the numerator and denominator of $I_i$ are now homogeneous of degree 4 in $\lam_{\hat{P}_i}$, we can simplify by multiplying both by $(\l<n1\r>[1 \hat{P}_i])^4$. Then every $\lam_{\hat{P}_i}$ now appears in the combination $\l<n1\r> [1 \hat{P}_i]\l< \hat{P}_i|...\,$. We can make use of the dual coordinates to remove all dependence on the hatted quantities as follows,
\begin{align}
&\l< n 1 \r>[1 \hat{P}_i] \l<\hat{P}_i|... = \l< n1 \r> [1|\hat{P}_i ...= \l< n1 \r> [1 | P_i ... \notag \\
&= \l<n1\r>[1|x_{1i}... = \l<n1\r> [1|x_{2i}... = \l<n|x_{12}x_{2i}... = \l<n|x_{n2}x_{2i}...\,.
\label{shuffling}
\end{align}
We can use these identities directly on the first term in the argument of the $\delta^4$ in (\ref{delta4}) so that it becomes
\be
-\l<n|x_{n2}x_{2i}|\theta_{i1}\r>.
\ee 
For the second term in the argument it is convenient to use the last form in the first line of (\ref{shuffling}) and to recall that the value of $z$ is fixed to be
\be
z= \frac{P_i^2}{\l<n|P_i|1]}.
\ee
This term then becomes
\be
- z \l< n1 \r> [1|P_i|n\r> \eta_1 = - P_i^2 \l< n1 \r> \eta_1 = -x_{1i}^2 \l<n 1\r> \eta_1 = \l< n|x_{1i}x_{i1}|1\r>\eta_1 = -\l<n|x_{ni} x_{i2} |\theta_{21}\r>.
\ee
Thus the $\delta^4$ factor becomes
\begin{align}
\delta^4 \bigl( \l<n|x_{n2}x_{2i}|\theta_{i1}\r> + \l<n|x_{ni}x_{i2}|\theta_{21}\r> \bigr) &= \delta^4 \bigl( \l<n|x_{n2}x_{2i}|\theta_{i}\r> + \l<n|x_{ni}x_{i2}|\theta_{2}\r>  + x_{2i}^2 \l<n \theta_1 \r>  \bigr)  
\label{rewritedelta4}
\end{align}
from which we can see that $\theta_1$ can be exchanged for $\theta_n$ as it comes projected with $\lam_n$.

Similar manipulations lead to the simplification of the bosonic factors in $I_i$. Finally we arrive at the result
\be
I_i = A_{n}^{\rm MHV} R_{n,2i},
\ee
where
\be
R_{n,2i}= \frac{\l<21\r> \l<i\,i-1\r> \delta^4\bigl( \l<n|x_{n2}x_{2i}|\theta_{in}\r> + \l<n|x_{ni}x_{i2}|\theta_{2n}\r> \bigr)}{x_{2i}^2 \l<n|x_{n2}x_{2i}|i\r> \l<n|x_{n2} x_{2i}|i-1\r> \l<n|x_{ni}x_{i2}|2\r> \l<n|x_{ni}x_{i2}|1\r>}.
\label{Rn2i}
\ee
Note that for the five-particle amplitude this is the only term in the amplitude because the first term in ({\ref{NMHVBCFW}) vanishes. For $n>5$ we can postulate the final form of the result and verify by induction that it is correct. The final form obtained for the NMHV amplitudes is
\be
\mathcal{A}_n^{\rm NMHV} = \mathcal{A}_n^{\rm MHV} \mathcal{P}_n^{\rm NMHV},
\label{ANMHV}
\ee
where
\be
\mathcal{P}_n^{\rm NMHV} = \sum_{2\leq a<b\leq n-1} \!\!\!\!\!\! R_{n,ab}.
\label{PNMHV}
\ee
Here $R_{n,ab}$ is the natural generalisation of (\ref{Rn2i}),
\be
R_{n,ab} = \frac{\l< a \, a-1\r> \l< b \, b-1\r> \delta^4 \bigl(\l< n |x_{na}x_{ab}|\theta_{bn}\r> + \l<n|x_{nb}x_{ba}|\theta_{an}\r>\bigr)}{x_{ab}^2 \l< n |x_{na} x_{ab} |b\r> \l< n|x_{na} x_{ab} |b-1\r> \l<n|x_{nb}x_{ba}|a\r> \l<n|x_{nb}x_{ba}|a\r> \l<n|x_{nb}x_{ba}|a-1\r>}
\label{Rnab}
\ee
and the sum over $a$ and $b$ in (\ref{PNMHV}) is performed such that $a<b-1$. The final result (\ref{PNMHV}) is remarkably simple. The six-particle case for example is expressed as sum of only three terms.

Note that in the result for the NMHV amplitudes, non-local poles make a prominent appearance. Poles of the form 
\be
\frac{1}{\l<n|x_{na}x_{ab}|b\r>}
\label{spurious}
\ee
are not (except for special values of $a$ and $b$) expressible in terms of the local poles of the type
\be
\frac{1}{(p_i + \ldots +p_{j})^2}.
\ee
Of course the theory is local and the spurious poles of the form (\ref{spurious}) cancel between different terms in the sum over the labels $a$ and $b$. However it is notable that the simple way of expressing the amplitude involves terms which are necessarily non-local. We will see a deeper reason for this in the next section.

The process of solving the recursion relation can be continued to higher levels in the MHV degree. We will not give the details of the calculations here but instead refer the reader to \cite{Drummond:2008cr} for the explicit derivations. The amplitudes are expressed in terms of new quantities that have many pairs of labels,
\be
R_{n;b_1a_1;b_2a_2;\ldots;b_ra_r;ab} = \frac{\l< a\,\,a-1\r> \l<
  b\,\,b-1\r> \delta^{(4)}(\l< \xi | x_{a_r a}x_{ab} | \theta_{ba_r} \r> +
  \l< \xi |x_{a_r b} x_{ba} | \theta_{aa_r} \r>)}{x_{ab}^2 \l< \xi |
  x_{a_r a} x_{ab} | b \r> \l< \xi | x_{a_r a} x_{ab} |b-1\r> \l< \xi
  | x_{a_r b} x_{ba} |a\r> \l< \xi | x_{a_r b} x_{ba} |a-1\r>}\,,
  \label{generalR}
\ee
where the chiral spinor $\langle \xi |$ is given by
\be
\l< \xi | = \l<n | x_{nb_1}x_{b_1 a_1} x_{a_1 b_2} x_{b_2 a_2} \ldots x_{b_r a_r} \,.
\ee
In the case where there is only one pair of labels $ab$ after the initial label $n$, (\ref{generalR}) is just the quantity $R_{n,ab}$ we have already seen appearing in the NMHV amplitudes. The cases where there is more than one pair are generalisations.

As an example we quote the result for NNMHV amplitudes here.
\begin{align}
\label{PNNMHVnew}
\mathcal{P}_n^{\rm NNMHV} = \sum_{2\leq a_1,b_1 \leq n-1}
\!\!\!\!\!\!\!\! R_{n;a_1b_1} \Biggl[ &\sum_{a_1+1 \leq a_2,b_2 \leq
  b_1} \!\!\!\!\!\!\!\! R_{n;b_1a_1;a_2b_2}^{0;a_1b_1}
  +  \sum_{b_1 \leq a_2b_2 \leq n-1} \!\!\!\!\!\!\!\! R_{n;a_2b_2}^{a_1b_1;0} \Biggr]\,.
\end{align}
We see that the generalised objects (\ref{generalR}) begin appearing at the NNMHV level in the first term in the brackets. The superscripts on the $R$s in the brackets mean that the boundary terms in the sum should be treated differently. The right superscript on $R_{n,b_1 a_1,a_2 b_2}$ indicates that when the upper boundary $b_2 = b_1$ is reached in the sum the explicit appearance of the spinor $\langle b_1 |$ should be replaced by $\langle n| x_{n a_1} x_{a_1 b_1}$. Similarly the left superscript on $R_{n,a_2 b_2}$ means that when the lower boundary $a_2=b_1$ is reached the explicit appearance of $\langle b_1 -1 |$ is replaced by $\langle n | x_{n a_1} x_{a_1 b_1}$. The 0 superscripts indicate that nothing happens at the other boundaries. The formulas for all N${}^p$MHV amplitudes can be found in \cite{Drummond:2008cr}.

As we have already discussed, at the level of pure gluon scattering, the fact that we have solved the amplitudes in the $\mathcal{N}=4$ theory is no restriction at all; the gluon amplitudes are the same in any gauge theory. Thus the simplicity of the expressions arising from the recursive structure is universal for gluon amplitudes in all gauge theories, as is the associated presence of spurious non-local poles. The explicit expressions for the pure-gluon amplitudes can be derived from (\ref{AMHV}) and (\ref{ANMHV},\ref{PNMHV},\ref{Rnab}) etc. by reading off the coefficient of the relevant combination of $\eta$ variables. 

The same recursive technique is also valid for gravitational theories \cite{Cachazo:2005ca}. Again it can be made manifestly supersymmetric and becomes much simpler for the maximally supersymmetric theory $\mathcal{N}=8$ supergravity \cite{ArkaniHamed:2008gz}, admitting an explicit solution \cite{Drummond:2009ge}.

\section{Dual superconformal symmetry}
\label{dualsconf}

Here we will see that the non-manifestly local form of the amplitudes arising from the solution of the BCFW recursion relation is very natural from the point of view of symmetry. Indeed the different terms arising in the BCFW expansion are all invariant under a very large symmetry algebra. 

Let us focus on the form of the MHV and NMHV amplitudes which we found in the previous section. The dual coordinates $x$ which are used to express the amplitudes can be taken seriously as the coordinates of a dual copy of spacetime. The amplitudes are trivially invariant under translations of the dual coordinates as they were introduced only through their differences. Lorentz transformations of the dual coordinates are the same as Lorentz transformations of the particle momenta and so are also a symmetry of the scattering amplitudes. The surprise comes when one examines conformal transformations of the dual coordinates. It turns out that these transformations are also a  symmetry of the scattering amplitudes. Since the symmetry acts canonically on the dual coordinates and these are linearly related to the particle momenta the generator is first order acting on the momenta. Note that such a conformal transformation is not related to the conformal symmetry of the Lagrangian, which is rather related to the second-order generators $k_{\a \adt}$ of (\ref{sconf}). The conformal symmetry acting in the dual space is referred to as dual conformal symmetry.

The dual coordinates $x_i$ define a closed polygon with light-like edges in the dual space, as represented in Fig. 11.
The contour is closed because we identify $x_{n+1}$ with $x_1$. This statement reflects the total momentum conservation of the scattering process $p_1+p_2+\ldots +p_n =0$. The edges of the polygon are light-like because the particles in the scattering process are on-shell. The role of the polygon was first made clear at strong coupling \cite{Alday:2007hr} where it becomes the boundary for a minimal surface in AdS space, leading to a relation between scattering amplitudes and Wilson loops. We will discuss this relation in more detail in the next section. For now we would just like to
note that a light-like polygon maps into another such polygon under conformal transformations of the dual space. Indeed under conformal inversions
\be
x^\mu \longrightarrow - \frac{x^\mu}{x^2},
\ee
we have that 
\be
x_{ij}^2 \longrightarrow \frac{x_{ij}^2}{x_i^2 x_j^2}.
\ee
Thus if two points $x_i$ and $x_j$ are light-like separated they will remain so after a conformal inversion. The conformal group is generated by Lorentz transformations, translations and conformal inversions so the light-like nature of the polygon is invariant under the action of the whole conformal group. 

\begin{figure}
\psfrag{x1}[cc][cc]{$x_1$}
\psfrag{x2}[cc][cc]{$x_2$}
\psfrag{x3}[cc][cc]{$x_3$}
\psfrag{x4}[cc][cc]{$x_4$}
\psfrag{xn}[cc][cc]{$x_n$}
\psfrag{p1}[cc][cc]{$p_1$}
\psfrag{p2}[cc][cc]{$p_2$}
\psfrag{p3}[cc][cc]{$p_3$}
\psfrag{pn}[cc][cc]{$p_n$}
 \centerline{{\epsfysize6cm
\epsfbox{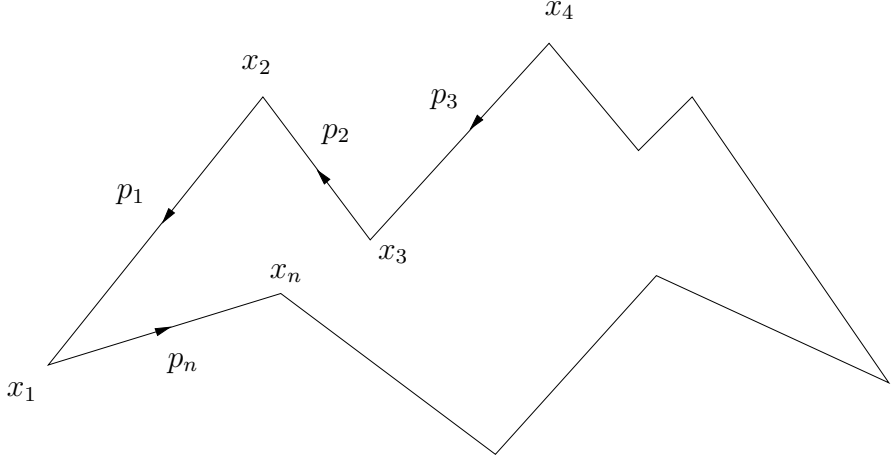}}}  \caption[]{\small The light-like polygon in dual coordinate space defined by the particle momenta.}
  \label{polygon}
\end{figure}

To see that dual conformal transformations are actually a symmetry of the scattering amplitudes we need to define their action on all of the variables in the problem. 
The helicity variables $\lam$ and $\tlam$ must also transform under dual conformal transformations as they are related to the dual coordinates via the defining relations,
\be
x_i^{\a \adt} - x_{i+1}^{\a \adt} = \lam_i^\a \tlam_i^{\adt}.
\label{dualx}
\ee
Indeed we see that $x_{ij}^{\a \adt}$ transforms as follows
\be
x_{ij}^{\a \adt} \longrightarrow -\frac{x_{i}^{\a \adt}}{x_i^2} + \frac{x_{j}^{\a \adt}}{x_j^2} = \frac{-x_i^{\a \adt}x_j^2 + x_i^2 x_j^{\a \adt}}{x_i^2 x_j^2} = \frac{ (x_i(x_i - x_j)x_j)^{\a \adt}}{x_i^2 x_j^2} = ( x_i^{-1}x_{ij}x_j^{-1} )^{\a \adt}.
\ee
Choosing $j=i+1$ we find
\be
\lam_i^\a \tlam_i^{\adt} \longrightarrow (x_i^{-1} \lam_i \tlam_i x_{i+1}^{-1} )^{\a \adt}.
\ee
The transformations $\lam_i$ and $\tlam_i$ can therefore be defined as follows \cite{Drummond:2008vq},
\be
\lam_i^{\a} \longrightarrow (x_i^{-1} \lam_i)^{\adt}, \qquad \tlam_i^{\adt} \longrightarrow (x_{i+1}^{-1} \tlam_i)^{\a},
\ee
so that they are compatible with the defining relations (\ref{dualx}).

The superpartners of the dual coordinates also transform canonically under conformal inversions,
\be
\theta_i^{\a A} \longrightarrow (x_i^{-1}\theta_i)^{\adt A},
\ee
which implies that the variables $\eta_i$ must also transform in analogy with $\lambda_i$ and $\tlam_i$. 
One can derive the transformation of the $\eta_i$ by performing an inversion on both sides of the relation
\be
\theta_i^{\a A} - \theta_{i+1}^{\a A} = \lambda_i^\a \eta_i^A.
\label{dualtheta}
\ee
It will not be necessary for us to give the transformation as we can always use (\ref{dualtheta}) to eliminate the $\eta_i$ in favour of the $\theta_i$ (we can similarly eliminate the $\tlam_i$ in favour of the $x_i$ using (\ref{dualx})). 

If we look at the MHV amplitude,
\be
\mathcal{A}_n^{\rm MHV} = \frac{\delta^4(p) \delta^8(q)}{\l< 12 \r> \ldots \l<n 1 \r>},
\ee 
we can see that it is in fact covariant under dual conformal transformations. Firstly, if we drop the requirement that $x_{n+1} \equiv x_1$ and $\theta_{n+1} \equiv \theta_1$ then the delta functions can be written as
\be
\delta^4(p) \delta^4(q) = \delta^4(x_1 - x_{n+1}) \delta^8(\theta_1 - \theta_{n+1}).
\label{deltas}
\ee 
This combination is dual conformally invariant as the bosonic delta function has conformal weight 4 at the point $x_1$ (which is identified with $x_{n+1}$ under the delta function) as can be seen from $\int d^4 x_1 \delta^4(x_1 - x_{n+1}) =1$. The Grassmann delta function has the opposite conformal weight (which is only true because of maximal supersymmetry) and the hence the product (\ref{deltas}) is invariant.

The denominator in the MHV amplitude is covariant under dual conformal transformations because factors of the form $\l< i \,i+1\r>$ transform as follows,
\be
\l<i \, i+1\r> \longrightarrow \l<i|x_i^{-1} x_{i+1}^{-1} |i+1\r> = \frac{\l< i|x_i x_{i+1} |i+1\r>}{x_i^2 x_{i+1}^2} = \frac{\l< i |x_{i+1} x_{i+1} |i+1 \r>}{x_i^2 x_{i+1}^2} = \frac{\l<i \, i+1\r>}{x_i^2}.
\label{abcovariance}
\ee
Thus we find that the MHV tree-level amplitude is covariant with weight 1 at each point,
\be
\mathcal{A}_n^{\rm MHV} \longrightarrow (x_1^2 \ldots x_n^2)\mathcal{A}_n^{\rm MHV}.
\ee

If we now look at the NMHV amplitude defined by equations (\ref{ANMHV}), (\ref{PNMHV}) and (\ref{Rnab}) we find that it is similarly covariant. The reason is that each term $R_{n,ab}$ in $\mathcal{P}_n^{\rm NMHV}$ is invariant under dual conformal transformations. Indeed returning to the formula (\ref{Rnab}) we see that it is made of dual conformally covariant factors. The spurious poles are covariant following a similar analysis to (\ref{abcovariance}).  For example we have
\be
\l< n|x_{na} x_{ab} |b \r> \longrightarrow \frac{\l<n|x_{na}x_{ab}|b\r>}{x_n^2 x_a^2 x_b^2}, \qquad \l<n|x_{na}x_{ab} |b-1\r> \longrightarrow \frac{\l<n| x_{na}x_{ab}|b-1\r>}{x_n^2 x_a^2 x_{b-1}^2}.
\ee
The Grassmann delta function is also covariant as can be see when written in a form similar to (\ref{rewritedelta4}),
\be
\delta^4 \bigl(\l< n |x_{na}x_{ab}|\theta_{bn}\r> + \l<n|x_{nb}x_{ba}|\theta_{an}\r>\bigr) = \delta^4 \bigl(\l< n |x_{na}x_{ab}|\theta_{b}\r> + \l<n|x_{nb}x_{ba}|\theta_{a}\r> + x_{ab}^2 \l< n \theta_n \r>\bigr)
\ee
Checking all the factors in (\ref{Rnab}) one can see that the weights cancel and thus $R_{n,ab}$ is invariant under dual conformal transformations. In fact one can show from the recursion relation itself that all terms produced this way will respect dual conformal symmetry \cite{Brandhuber:2008pf}. One can also check directly that dual conformal symmetry is present for the generalisations of $R_{n,ab}$ appearing in the N$^p$MHV amplitudes.

The dual conformal symmetry we have seen actually extends to dual superconformal symmetry. Dual superconformal symmetry has a canonical action on the coordinates of the dual superspace $x_i,\theta_i$.
It is very helpful to express the symmetry in terms of the generators of infinitesimal transformations. For example the generator of special conformal transformations of the dual coordinates is
\be
K_{\alpha \dot{\alpha}} = \sum_i [x_{i \alpha}{}^{\dot{\beta}} x_{i
    \dot{\alpha}}{}^{\beta} \partial_{i \beta \dot{\beta}} + x_{i
    \dot{\alpha}}{}^{\beta} \theta_{i \alpha}^B \partial_{i \beta B}].
\label{dualK}
\ee
Just as we have seen for dual conformal inversions the transformation must act on the on-shell superspace variables $\{\lam,\tlam,\eta\}$ in order to respect the relations between them (constraints). In terms of the generators this means we must add terms so that the generator commutes with the constraints modulo constraints. One can perform this process for all generators of the superconformal algebra. The result is the following set of generators,
\begin{align}
P_{\alpha \dot{\alpha}}&= \sum_i \partial_{i \alpha \dot{\alpha}}\,, \qquad Q_{\alpha A} = \sum_i \partial_{i \alpha A}\,, \qquad
\overline{Q}_{\dot{\alpha}}^A = \sum_i [\theta_i^{\alpha A}
  \partial_{i \alpha \dot{\alpha}} + \eta_i^A \partial_{i \dot{\alpha}}], \notag\\
M_{\alpha \beta} &= \sum_i[x_{i ( \alpha}{}^{\dot{\alpha}}
  \partial_{i \beta ) \dot{\alpha}} + \theta_{i (\alpha}^A \partial_{i
  \beta) A} + \lambda_{i (\alpha} \partial_{i \beta)}]\,, \qquad
\overline{M}_{\dot{\alpha} \dot{\beta}} = \sum_i [x_{i
    (\dot{\alpha}}{}^{\alpha} \partial_{i \dot{\beta} ) \alpha} +
  \tilde{\lambda}_{i(\dot{\alpha}} \partial_{i \dot{\beta})}]\,,\notag\\
R^{A}{}_{B} &= \sum_i [\theta_i^{\alpha A} \partial_{i \alpha B} +
  \eta_i^A \partial_{i B} - \tfrac{1}{4} \delta^A_B \theta_i^{\alpha
    C} \partial_{i \alpha C} - \tfrac{1}{4}\delta^A_B \eta_i^C \partial_{i C}
]\,,\notag\\
D &= \sum_i [-x_i^{\dot{\alpha}\alpha}\partial_{i \alpha \dot{\alpha}} -
  \tfrac{1}{2} \theta_i^{\alpha A} \partial_{i \alpha A} -
  \tfrac{1}{2} \lambda_i^{\alpha} \partial_{i \alpha} -\tfrac{1}{2}
  \tilde{\lambda}_i^{\dot{\alpha}} \partial_{i \dot{\alpha}}]\,,\notag\\
C &=  \sum_i [-\tfrac{1}{2}\lambda_i^{\alpha} \partial_{i \alpha} +
  \tfrac{1}{2}\tilde{\lambda}_i^{\dot{\alpha}} \partial_{i \dot{\alpha}} + \tfrac{1}{2}\eta_i^A
  \partial_{i A}]\,, \notag\\
S_{\alpha}^A &= \sum_i [-\theta_{i \alpha}^{B} \theta_i^{\beta A}
  \partial_{i \beta B} + x_{i \alpha}{}^{\dot{\beta}} \theta_i^{\beta
    A} \partial_{i \beta \dot{\beta}} + \lambda_{i \alpha}
  \theta_{i}^{\gamma A} \partial_{i \gamma} + x_{i+1\,
    \alpha}{}^{\dot{\beta}} \eta_i^A \partial_{i \dot{\beta}} -
  \theta_{i+1\, \alpha}^B \eta_i^A \partial_{i B}]\,,\notag\\
\overline{S}_{\dot{\alpha} A} &= \sum_i [x_{i \dot{\alpha}}{}^{\beta}
  \partial_{i \beta A} + \tilde{\lambda}_{i \dot{\alpha}}
  \partial_{iA}]\,,\notag\\
K_{\alpha \dot{\alpha}} &= \sum_i [x_{i \alpha}{}^{\dot{\beta}} x_{i
    \dot{\alpha}}{}^{\beta} \partial_{i \beta \dot{\beta}} + x_{i
    \dot{\alpha}}{}^{\beta} \theta_{i \alpha}^B \partial_{i \beta B} +
  x_{i \dot{\alpha}}{}^{\beta} \lambda_{i \alpha} \partial_{i \beta}
  + x_{i+1 \,\alpha}{}^{\dot{\beta}} \tilde{\lambda}_{i \dot{\alpha}}
  \partial_{i \dot{\beta}} + \tilde{\lambda}_{i \dot{\alpha}} \theta_{i+1\,
    \alpha}^B \partial_{i B}]\,.
\label{dualsc}
\end{align}
Here we have employed the following shorthand notation
\begin{align}\label{shortderiv}
\partial_{i \alpha \dot{\alpha}} = \frac{\partial}{\partial
x_i^{\alpha \dot{\alpha}}}, \qquad \partial_{i \alpha A} = \frac{\partial}{\partial \theta_i^{\alpha
A}}, \qquad \partial_{i \alpha} = \frac{\partial}{\partial \lambda_i^{\alpha}}\,, \qquad
\partial_{i \dot{\alpha}} = \frac{\partial}{\partial
    \tilde{\lambda}_i^{\dot{\alpha}}}\,, \qquad
\partial_{i A} = \frac{\partial}{\partial \eta_i^A}\,.
\end{align}

It is simple to check that the generators in (\ref{dualsc}) do obey the commutation relations of the superconformal algebra. There are several remarks worth making at this point. Firstly it is clear from the first-order form of the generators that the dual superconformal symmetry is distinct from the ordinary superconformal symmetry generated by (\ref{superPoincaremulti}),(\ref{sconf}) etc. Secondly, we note that the $su(4)$ nature of the fermionic generators is swapped between the original superconformal algebra and the dual superconformal algebra. For example the supersymmetry generator $q^{\a A}$ is in the fundamental representation of $su(4)$ while the dual supersymmetry generator $Q_{\a A}$ is in the anti-fundamental. Similarly, on the on-shell superspace variables, the two dilatation generators coincide up to a sign because dual coordinates are actually related to particle momenta. Finally we should note that the two superconformal algebras overlap non-trivially. That is, the fermionic superconformal generator $\bar{S}$ coincides with the original supersymmetry generator $\bar{q}$ on the on-shell superspace, while the dual supersymmetry generator $\bar{Q}$ coincides with the original superconformal generator $\bar{s}$. The definitions of the dual variables manifestly respect covariance under the Lorentz, dilatation and $su(4)$ symmetries and so these symmetries are shared between the two copies of the superconformal algebra. The overlap is schematically represented in the following picture.

\vspace{10mm} \psfrag{p}[cc][cc]{\parbox[t]{0mm}{ {\Large$p$}}}
\psfrag{q}[cc][cc]{\parbox[t]{0mm}{
{\Large$q$}}} \psfrag{s}[cc][cc]{\parbox[t]{0mm}{ {\Large$s$}}}
\psfrag{k}[cc][cc]{\parbox[t]{0mm}{
{\Large$k$}}}
\psfrag{P}[cc][cc]{\parbox[t]{0mm}{ {\Large$P$}}}
\psfrag{K}[cc][cc]{\parbox[t]{0mm}{ {\Large$K$}}}
\psfrag{S}[cc][cc]{\parbox[t]{0mm}{ {\Large$S$}}}
\psfrag{Q}[cc][cc]{\parbox[t]{0mm}{ {\Large$Q$}}}
\psfrag{bq}[cc][cc]{\Large $\bar{q} \,=
\bar{S}$ }
\psfrag{bs}[cc][cc]{\Large  $\bar{s} \,=
\bar{Q}$}
\centerline{{\epsfysize4.5cm \epsfbox{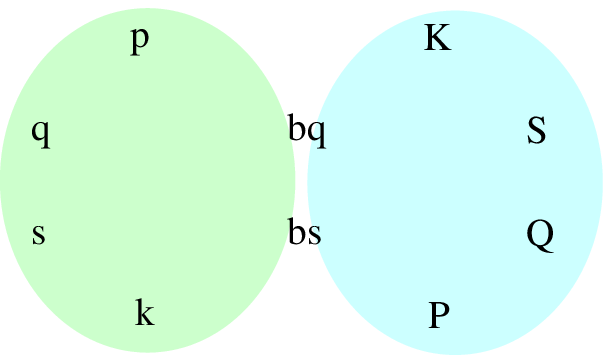}}} \vspace{0mm}
\noindent 
A similar picture also arises from considering the symmetries of the string sigma model (see \cite{Ricci:2007eq,Berkovits:2008ic,Beisert:2008iq}) which can be used to describe scattering amplitudes at strong coupling.

With all the generators of the superconformal algebra to hand we can now verify that the quantity $R_{n,ab}$ is actually a dual superconformal invariant. We have already verified that it is invariant under dual conformal inversions. Since dual translation invariance and Lorentz invariance are manifest the inversion symmetry is equivalent to invariance under the dual special conformal transformations, generated by $K_{\a \adt}$. 
It remains to show that it is invariant under the fermionic generators. Invariance under the chiral dual supersymmetry $Q_{\a A}$ is manifest (the $\theta$ variables only appear as differences in $R_{n,ab}$) and hence by commutation with $K_{\a \adt}$ we know that $\bar{S}_{\adt A}$ is a symmetry. The non-trivial symmetry to verify is the anti-chiral dual supersymmetry $\bar{Q}_\adt^A$.

To show that $\bar{Q}_{\adt}^{A}$ is indeed a symmetry of $R_{n,ab}$ we can exploit $Q$ and $\bar{S}$ by using a finite transformation made from these generators to fix a frame where $\theta_a=\theta_b=0$. Since $R_{n,ab}$ is invariant under $Q$ and $\bar{S}$ and all generators which arise through commutation of these with $\bar{Q}$ we know that $\bar{Q} R_{n,ab}$ is invariant under $Q$ and $\bar{S}$. So we can evaluate $\bar{Q} R_{n,ab}$ in the frame where $\theta_a=\theta_b=0$,
\begin{align}
\bar{Q}_{\adt}^{A} R_{n,ab} &= \theta_{n}^{\a A} \frac{\partial}
{\partial x_{n}^{\a \adt} } \biggl( \frac{ \l< a \, a-1\r> \l< b \, b-1\r> \delta^4\bigl( \l< n \theta_n \r> \bigr) }{ x_{ab}^2 \l< n | x_{na} x_{ab} | b \r> \l< n |x_{na} x_{ab} | b-1\r> \l<n|x_{nb}x_{ba} | a \r> \l< n |x_{nb} x_{ba} |a-1\r>}
 \biggr). \notag \\
 & \propto \l< n \theta_n \r> \delta^4\bigl( \l< n \theta_n \r>\bigr) = 0.
\end{align}
Thus we can see that the nilpotent nature of $R_{n,ab}$ is crucial in satisfying the invariance.

As we have seen the higher $R$-invariants appearing in the tree-level S-matrix are also dual conformal invariants. They are not dual superconformal invariants, as they are not annihilated by $\bar{Q}$. However they always appear in a nested fashion. For example at NNMHV level the quantity $R_{n,b_1 a_1, a_2 b_2}$ always appears multiplied by $R_{n,a_1 b_1}$. The $\bar{Q}$ variation of $R_{n;b_1 a_1,a_2 b_2}$ vanishes when multiplied by $R_{n,a_1b_1}$ so that the product is again dual superconformal invariant. Again we refer the reader to \cite{Drummond:2008cr} for more details.

The end result for the full superamplitude is that the function $\mathcal{P}_n$ is invariant under dual superconformal symmetry while the MHV prefactor is covariant under $D$, $C$, $K$ and $S$ and invariant under all other dual superconformal transformations. Thus we have
\be
D \mathcal{A}_n = n \mathcal{A}_n, \qquad C \mathcal{A}_n = n \mathcal{A}_n, \qquad  K^{\a \adt} \mathcal{A}_n = -\sum_i x_i^{\a \adt} \mathcal{A}_n, \qquad S^{\a A} \mathcal{A}_n = -\sum_i \theta_i^{\a A} \mathcal{A}_n
\label{dualsconfcov}
\ee
In addition we have invariance under the standard superconformal symmetry (see \cite{Witten:2003nn,ArkaniHamed:2009si,Mason:2009sa,Korchemsky:2009jv} for a fuller discussion),
\be
j_a \mathcal{A}_n = 0.
\label{scs}
\ee

We can get some insight into the nature of the symmetries we have found by combining the dual superconformal symmetry with the original one. In order to put the dual superconformal symmetry on the same footing as invariance under the standard superconformal algebra (\ref{scs}), the covariance (\ref{dualsconfcov}) can be rephrased as an invariance of $\mathcal{A}_n$ by a simple redefinition of the generators \cite{Drummond:2009fd},
\begin{align}
K'^{\alpha \dot{\alpha}} &= K^{\alpha \dot{\alpha}} + \sum_i x_i^{\alpha \dot{\alpha}}, \\
S'^{\alpha A} &= S^{\alpha A} + \sum_i \theta_i^{\alpha A}, \\
D' &= D - n.
\end{align}
With this redefinition all generators of the dual superconformal algebra annihilate $\mathcal{A}_n$.

In order to have both symmetries acting on the same space it is useful to restrict the dual superconformal generators to act only on the on-shell superspace variables $(\lambda_i,\tilde{\lambda}_i,\eta_i)$. Then one finds that the generators $P_{\alpha \dot\alpha},Q_{\alpha A}$ become trivial while the generators $\{\bar{Q},M,\bar{M},R,D',\bar{S}\}$ coincide (up to signs) with generators of the standard superconformal symmetry. The non-trivial generators which are not part of the $j_a$ are $K'$ and $S'$. In \cite{Drummond:2009fd} it was shown that the generators $j_a$ and $S'$ (or $K'$) together generate the Yangian of the superconformal algebra, $Y(psu(2,2|4))$. The generators $j_a$ form the level-zero $psu(2,2|4)$ subalgebra\footnote{We use the symbol $[O_1,O_2]$ to denote the bracket of the Lie superalgebra, $[O_2,O_1] = (-1)^{1+|O_1||O_2|}[O_1,O_2]$.},
\begin{equation}
[j_a,j_b] = f_{ab}{}^{c} j_c.
\end{equation}
In addition there are level-one generators $j_a^{(1)}$ which transform in the adjoint under the level-zero generators,
\begin{equation}
[j_a,j_b\!{}^{(1)}] = f_{ab}{}^{c} j_c\!{}^{(1)}.
\end{equation}
Higher commutators among the generators are constrained by the Serre relation\footnote{The symbol $\{\cdot,\cdot,\cdot\}$ denotes the graded symmetriser.},
\begin{align}
&[j^{(1)}_a , [j^{(1)}_b,j_c]] + (-1)^{|a|(|b| + |c|)} [j^{(1)}_b,[j^{(1)}_c,j_a]] + (-1)^{|c|(|a|+|b|)} [j^{(1)}_c,[j^{(1)}_a,j_b]] \notag \\
&= h^2 (-1)^{|r||m|+|t||n|}\{j_l,j_m,j_n\} f_{ar}{}^{l} f_{bs}{}^{m} f_{ct}{}^{n} f^{rst}.
\end{align}
The level-zero generators are represented by a sum over single particle generators,
\begin{equation}
j_a = \sum_{k=1}^n j_{ka}.
\label{levelzero}
\end{equation}
The level-one generators are represented by the bilocal formula \cite{Dolan:2003uh,Dolan:2004ps},
\begin{equation}
j_a\!{}^{(1)} = f_{a}{}^{cb} \sum_{k<k'} j_{kb} j_{k'c}.
\label{bilocal}
\end{equation}
Thus finally the full symmetry of the tree-level amplitudes can be rephrased as
\begin{equation}
y \mathcal{A}_n = 0,
\end{equation}
for any $y \in Y(psu(2,2|4))$.

One can naturally describe the symmetry in terms of twistor variables. In $(2,2)$ signature the twistor variables are simply related to the on-shell superspace variables $(\lambda,\tilde{\lambda},\eta)$ by a Fourier transformation $\lambda \longrightarrow \tilde{\mu}$ \cite{Witten:2003nn}. Twistor space linearises the action of superconformal symmetry.
Expressed in terms of the twistor space variables $\mathcal{Z}^{\AA} = (\tilde{\mu}^{\alpha}, \tilde\lambda^{\dot\alpha} , \eta^A)$, the level-zero and level-one generators of the Yangian symmetry assume a simple form
\begin{align}
j^{\AA}{}_{\BB} &= \sum_i \cZ_i^{\AA} \frac{\partial}{\partial \cZ_i^{\BB}}, \label{twistorsconf}\\
j^{(1)}{}^{\AA}{}_{\BB} &= \sum_{i<j} (-1)^{\CC}\Bigl[\cZ_i^{\AA} \frac{\partial}{\partial \cZ_i^{\CC}} \cZ_j^{\CC} \frac{\partial}{\partial \cZ_j^{\BB}} - (i,j) \Bigr].
\label{twistoryangian}
\end{align}
Both of the formulas (\ref{twistorsconf}) and (\ref{twistoryangian}) are understood to have the supertrace 
removed. In this representation the generators of superconformal symmetry are first-order operators while the level-one Yangian generators are second order.

In fact one can also phrase things the other way round. In \cite{Drummond:2010qh} it was demonstrated that there exists an alternative T-dual representation of the symmetry, where it is the dual superconformal generators $J_a$ which play the role of the level-zero generators, while the generators $k$ and $s$ of the original superconformal symmetry form part of the level-one generators. The generators take the same form as (\ref{twistorsconf}) and (\ref{twistoryangian}) but with the twistor variables replaced by momentum twistor variables \cite{Hodges:2009hk}. Momentum twistors are the twistors associated to the dual space and linearise the dual superconformal symmetry. They are defined as $\mathcal{W}_i^{\AA} = (\lambda_i^\a,\mu_i^\adt,\chi_i^A)$ with
\be
\mu_i^\a = x^{\a \adt}_i \lam_{i\a}, \qquad \chi_i^A = \theta^{\a A}_i \lam_{i \a}.
\ee
The generators then take the form
\begin{align}
J^{\AA}{}_{\BB} &= \sum_i \cW_i^{\AA} \frac{\partial}{\partial \cW_i^{\BB}}, \label{momtwistordsconf}\\
J^{(1)}{}^{\AA}{}_{\BB} &= \sum_{i<j} (-1)^{\CC}\Bigl[\cW_i^{\AA} \frac{\partial}{\partial \cW_i^{\CC}} \cW_j^{\CC} \frac{\partial}{\partial \cW_j^{\BB}} - (i,j) \Bigr],
\label{momtwistoryangian}
\end{align}
again with supertraces removed. In this representation the generators annihilate the amplitude with the MHV prefactor dropped, i.e.
\be
J_a \mathcal{P}_n = 0, \qquad J^{(1)}_a \mathcal{P}_n = 0.
\ee

So we have seen that there are two equivalent ways of looking at the full symmetry of theory. One can choose either version of the superconformal symmetry to be fundamental. This effectively means choosing one of the superconformal symmetries to be realised locally. In so doing the remaining symmetries are necessarily non-local. Thus the non-local poles found in the expression for the tree-level amplitudes are inevitable if the amplitudes are to be expressed in a form which reveals the full symmetry.

The fact that the full symmetry is the Yangian $Y(psu(2,2|4))$ is certainly not accidental. The planar limit of the $\mathcal{N}=4$ gauge theory is known to possess an integrable structure in other regimes. In particular the believed integrability of the spectrum of anomalous dimensions (see e.g. \cite{Minahan:2002ve,Beisert:2005fw,Beisert:2006ez,Gromov:2009tv,Arutyunov:2009ur}) can be traced to the fact that there is an underlying Yangian symmetry. Moreover at strong coupling the theory is related via the AdS/CFT correspondence to the $AdS_5$ sigma model which is classically integrable \cite{Bena:2003wd}. Indeed the integrability of this model has been used to calculate the scattering amplitudes via a relation to minimal surfaces in AdS which we will describe in the next section.

Having described the symmetry of the theory, one might naturally ask how one can produce invariants. This question has been addressed in various papers \cite{Korchemsky:2009jv,Drummond:2010qh,Drummond:2010uq,Korchemsky:2010ut}. It turns out to be intimately connected to another conjecture about the leading singularities of the scattering amplitudes of $\mathcal{N}=4$ super Yang-Mills theory. In \cite{ArkaniHamed:2009dn} Arkani-Hamed et al proposed a formula for all leading singularities of the planar $\mathcal{N}=4$ S-matrix. A leading singularity of a loop amplitude is obtained by evaluating the loop integration via compact contours instead of the usual non-compact Minkowski space integration. An example is given by the four-particle cuts of \cite{Britto:2004nc}. Here one takes the one-loop amplitude and evaluates it by choosing a contour which localises the loop integration to a point where four internal propagators go on-shell. There are four parameters in the loop integration variable which are fixed by choosing four constraints to be satisfied. There are in general two solutions to these conditions, each of which is a leading singularity. See \cite{Cachazo:2008vp,Cachazo:2008hp,ArkaniHamed:2009dn} for more discussions of leading singularities.
 
The formula of \cite{ArkaniHamed:2009dn} takes the form of an integral over the Grassmannian $G(k,n)$, where $k$ is the level in the MHV expansion and the $n$ is the number of particles. It can be expressed either in the original twistor space \cite{ArkaniHamed:2009dn} or in momentum twistor space \cite{Mason:2009qx}.
The formula possesses one manifest superconformal symmetry and one non-manifest one and hence it produces (for different choices of the contour of integration) different Yangian invariants, including the ones we have seen appearing from the BCFW expansion. The twistor and momentum twistor versions of the formula can be directly related \cite{ArkaniHamed:2009vw} showing that both versions do possess both copies of the superconformal symmetry.
In fact this formula is the most general way of producing such an invariant \cite{Drummond:2010uq,Korchemsky:2010ut}. If the conjecture of \cite{ArkaniHamed:2009dn} is correct then we see that the Yangian symmetry also plays a role at loop level by constraining the form of the leading singularities. In the next section we will see how at least part of the symmetry also constrains the form of the loop amplitudes themselves and not just the leading singularities.

\section{Loop corrections}
\label{loops}

Having examined in detail the structure of the amplitudes and symmetries at tree-level it is natural to ask what happens when perturbative corrections are taken into account. In this section we will review some of the features of scattering amplitudes in perturbation theory. As our main motivation is to understand the extended symmetries we discussed in the previous section, we will be concerned entirely with planar amplitudes in the $\mathcal{N}=4$ theory. 
Many of the developments we have already outlined at tree-level were in fact preceded by various observations for loop corrections to scattering amplitudes. As we will see the dual conformal symmetry can be directly observed in the form of the loop integrals appearing in the planar scattering amplitudes of $\mathcal{N}=4$ super Yang-Mills theory. A lot can be learned from the simplest case, namely the four-gluon scattering amplitudes. As we have seen, these amplitudes are examples of the so-called maximally-helicity-violating or MHV amplitudes.

MHV amplitudes are particularly simple in that they can naturally be written as a product of the rational tree-level amplitude and a loop-correction function which is a series in the `t Hooft coupling $a$,
\begin{equation}
\mathcal{A}_n^{\rm MHV} = \mathcal{A}_{n,{\rm tree}}^{\rm MHV} \, M_n(p_1,\ldots,p_n;a).
\end{equation}
By parity we could of course equivalently study the $\overline{\rm MHV}$ amplitudes\footnote{As we saw already, the four-particle amplitudes are actually both MHV and $\overline{\rm MHV}$.}

In perturbation theory the function $M_n$ is expressed in terms of scalar loop integrals. 
Let us consider the four-gluon amplitude at one loop. The only contribution to $M_4$ at this order is given by the scalar box integral,
\begin{equation}\label{1box}
    I^{(1)} = \int \frac{d^4k}{k^2(k-p_1)^2(k-p_1-p_2)^2(k+p_4)^2}\ .
\end{equation}
This integral formally exhibits dual conformal symmetry. To make this apparent we will employ our usual change of variables from momenta to dual coordinates,
\begin{equation}
p_i^\mu = x_i^\mu - x_{i+1}^\mu \equiv x_{i,i+1}^\mu,
\label{dualcoords}
\end{equation}
with $x_5 \equiv x_1$.
After the change of variables (\ref{dualcoords}) the integral can then be written as a four-point star diagram (the dual graph for the one-loop box) with the loop integration replaced by an integration over the internal vertex $x_5$ (see  Fig. \ref{figure:dualdiag}).
\begin{figure}[htbp]
\psfrag{x1}[cc][cc]{$x_1$} \psfrag{x2}[cc][cc]{$x_2$} \psfrag{x3}[cc][cc]{$x_3$}
\psfrag{x4}[cc][cc]{$x_4$} \psfrag{p1}[cc][cc]{$p_1$} \psfrag{p2}[cc][cc]{$p_2$}
\psfrag{p3}[cc][cc]{$p_3$} \psfrag{p4}[cc][cc]{$p_4$} \psfrag{x5}[cc][cc]{$x_5$}
\centerline{{\epsfysize5cm \epsfbox{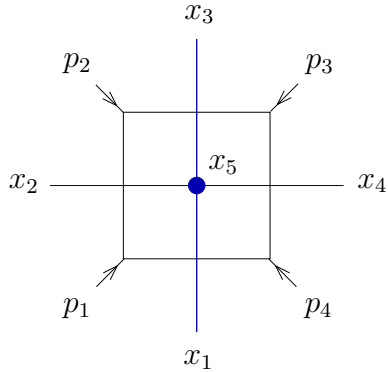}}} \caption{\small  Dual diagram
for the one-loop box.}
\label{figure:dualdiag}
\end{figure}

If we consider conformal inversions of the dual coordinates,
\begin{equation}
x_i^\mu \longrightarrow - \frac{x_i^{\mu}}{x_i^2},
\end{equation}
then we see that the integrand, including the measure factor $d^4x_5$, is covariant.
\begin{equation}
\frac{d^4 x_5}{x_{15}^2 x_{25}^2 x_{35}^2 x_{45}^2} \longrightarrow (x_1^2 x_2^2 x_3^2 x_4^2) \frac{d^4 x_5}{x_{15}^2 x_{25}^2 x_{35}^2 x_{45}^2}.
\end{equation}
If we had normalised the integral with an extra factor of $x_{13}^2 x_{24}^2$ then it would actually be invariant.

The property of dual conformal covariance of the integral form is not restricted to one loop but continues to all loop orders so far explored \cite{Bern:2006ew,Bern:2007ct}. In Fig. \ref{figure:4ptamp} we show the integral topologies occurring in the four-point amplitude up to three loops.
\begin{figure}[htbp]
\centerline{{\epsfysize5cm \epsfbox{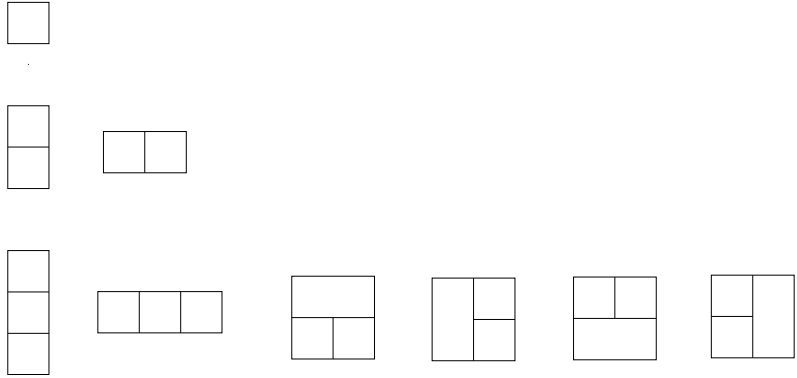}}}
\caption{\small Integral topologies appearing in the four particle amplitude up to three loops.}
\label{figure:4ptamp}
\end{figure}
All of these integrals exhibit dual conformal symmetry in the same sense as the one-loop scalar box. 

The dual conformal property is rather restrictive in the kinds of integrals it allows to appear. At two loops there is again only one topology, the two-loop scalar box. At three loops there are two topologies, the three-loop box and the so-called `tennis court'.  The tennis court requires a precise numerator factor to be dual conformally covariant (see Fig. \ref{figure:3ladandtc}).
\begin{figure}[htbp]
\psfrag{x1}[cc][cc]{\,\,\,\,$x_1$}
\psfrag{x2}[cc][cc]{$x_2$}
\psfrag{x3}[cc][cc]{\,\,\,\,$x_3$}
\psfrag{x4}[cc][cc]{$x_4$}
\psfrag{x5}[cc][cc]{$x_5$}
\psfrag{x6}[cc][cc]{$x_6$}
\psfrag{x7}[cc][cc]{$x_7$}
\psfrag{p1}[cc][cc]{$p_1$}
\psfrag{p2}[cc][cc]{$p_2$}
\psfrag{p3}[cc][cc]{$p_3$}
\psfrag{p4}[cc][cc]{$p_4$}
\centerline{{\epsfysize6.5cm \epsfbox{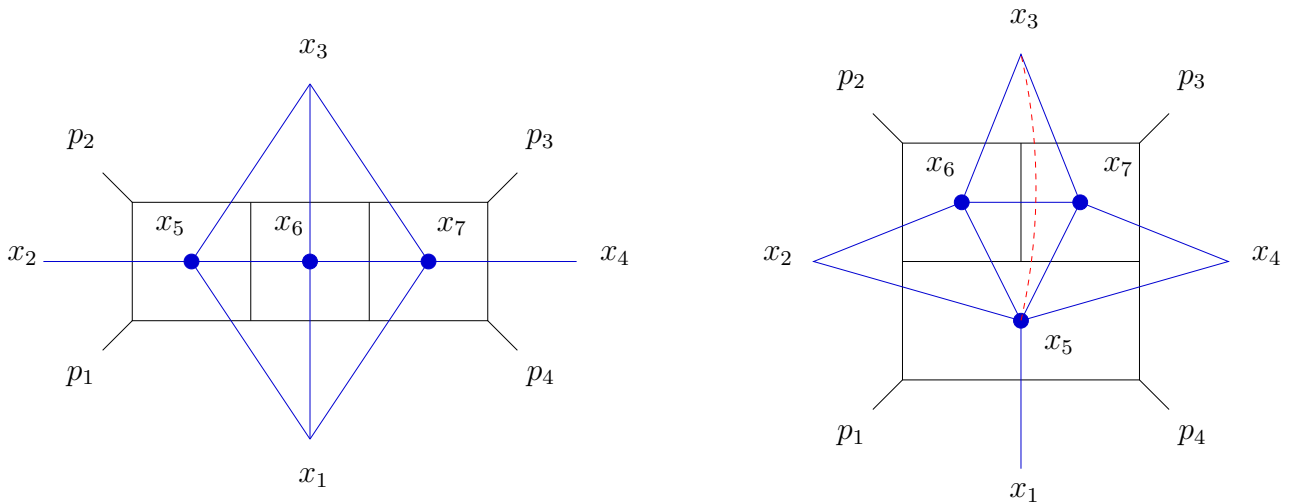}}}
\caption{\small Dual diagrams for the three-loop box and for the `tennis court'
with its numerator denoted by the dashed line corresponding to a factor $x_{35}^2$.} \label{figure:3ladandtc}
\end{figure}
Note that the operation of drawing the dual graph is only possible for planar diagrams. This fits with our expectation that the symmetry is related to the integrable structure of the planar theory.

The integrals, e.g. as defined in (\ref{1box}), are infrared divergent. This is to be expected as amplitudes in massless theories inevitable exhibit infrared divergences. One therefore need to consistently introduce a regulator in order to talk about the S-matrix beyond tree level. A common choice is to use dimensional regularisation, by taking the original Lagrangian in $4-2\epsilon$ dimensions, with $\epsilon < 0$ regulating the infrared divergences. This breaks the dual conformal symmetry slightly since the measure is then no longer in four dimensions,
\begin{equation}
d^4x_5 \longrightarrow d^{4-2\epsilon} x_5.
\end{equation}
Another choice for dealing with the infrared divergences is to study the theory on the Coulomb branch \cite{Alday:2009zm}. Now the VEVs of the scalar fields  regulate the infrared divergences by introducing masses in a particular way for the virtual particles propagating in the loops. One can think of this regularisation geometrically; the dual Minkowski space is the boundary of a five-dimensional $AdS$ space and one moves the points $x_1,x_2,x_3$ and $x_4$ slightly off the boundary into the bulk. The masses are interpreted as radial coordinates in the $AdS_5$ space and the corresponding action of dual conformal symmetry leaves the correctly normalised integral invariant\footnote{In addition to this conceptual advantage this regularisation also has practical advantages, see \cite{Henn:2010bk,Henn:2010ir} for more discussion.}. One can therefore say that it is a function of the invariants one can construct from the dual $x_i$ and the masses (which also transform under the action of the dual conformal symmetry).

With either choice of regularisation the consequences of the new symmetry as the regulators are taken to zero are not immediately apparent. In dimensional regularisation one needs to know more precisely how the symmetry is broken by the $O(\epsilon)$ effects. In the $AdS_5$ regularisation one needs to know precisely how the amplitudes are allowed to depend on the radial coordinates (or masses) in the limit in which they become small. 

Fortunately, there is a dual picture which allows us to understand the breaking of dual conformal symmetry in a precise way and also potentially sheds light on its origin. In the dual description planar MHV amplitudes are related to Wilson loops defined on a piecewise light-like contour in the dual coordinate space. The contour is none other than the light-like polygon in Fig. \ref{polygon} that we have seen arising from the definition of the dual coordinates. In a gauge theory it is very natural to associate a Wilson loop to this contour, 
\begin{equation}
W_n = \langle \mathcal{P}exp \oint_{C_n} A \rangle.
\label{Wilsonloop}
\end{equation}
Here, in contrast to the situation for the scattering amplitude, the dual space is being treated as the actual configuration space of the gauge theory, i.e. the theory in which we compute the Wilson loop is local in this space.

So let us consider the general structure of MHV amplitudes and light-like Wilson loops in $\mathcal{N}=4$ super Yang-Mills theory. We will begin with the MHV amplitude. As we have discussed we can naturally factorise MHV amplitudes into a tree-level factor and a loop-correction factor $M_n$. The factor $M_n$ is infrared divergent and will therefore depend on the regularisation parameters. We will use dimensional regularisation so $M_n$ depends on the regulator $\epsilon$ and an associated scale $\mu$.

Since we are discussing planar colour-ordered amplitudes it is clear that the infrared divergences will involve only a very limited dependence on the kinematical variables. Specifically, the exchange of soft or collinear gluons is limited to sectors between two adjacent incoming particles and thus the infrared divergences will factorise into pieces which depend only on a single Mandelstam variable $s_{i,i+1}=(p_i + p_{i+1})^2$.
Moreover the dependence of each of these factors is known to be of a particular exponentiated form \cite{Akhoury:1978vq,Mueller:1979ih,Collins:1980ih,Sen:1981sd,Sterman:1986aj,Botts:1989kf,Catani:1989ne,Korchemsky:1988hd,Korchemsky:1988pn,Magnea:1990zb,Korchemsky:1993uz,Catani:1998bh,Sterman:2002qn} where there is at most a double pole in the regulator in the exponent. It is therefore natural to consider the logarithm of the loop corrections $M_n$,
\begin{equation}
{\rm log}\,M_n = \sum_{l=1}^{\infty} a^l \biggl[ \frac{\Gamma_{\rm cusp}^{(l)}}{(l\epsilon)^2} + \frac{\Gamma_{\rm col}^{(l)}}{l \epsilon} \biggr] \sum_i\biggl(\frac{\mu_{IR}^2}{-s_{i,i+1}}\biggr)^{l\epsilon} + F^{\rm MHV}_n(p_1,\ldots,p_n;a) + O(\epsilon).
\label{logMn}
\end{equation}
The leading infrared divergence is known to be governed by $\Gamma_{\rm cusp}(a) = \sum a^l \Gamma_{\rm cusp}^{(l)}$, the cusp anomalous dimension \cite{Ivanov:1985np,Korchemsky:1992xv}, a quantity which is so called because it arises as the leading ultraviolet divergence of Wilson loops with light-like cusps. 
This is a first hint at the connection between scattering amplitudes and Wilson loops. 

In \cite{Bern:2005iz} Bern-Dixon and Smirnov (BDS) made an all order ansatz for the form of the finite part of the $n$-point MHV scattering amplitude. Their ansatz had the following form,
\begin{equation}
F_n^{\rm BDS}(p_1,\ldots,p_n;a) = \Gamma_{\rm cusp}(a) \mathcal{F}_n(p_1,\ldots,p_n) + c_n(a).
\label{BDSansatz}
\end{equation}
The notable feature of this ansatz is that the dependence on the coupling arises only via a single function, the cusp anomalous dimension, while the momentum dependence is contained in the coupling-independent function $\mathcal{F}_n$. The latter could therefore be defined by the one-loop amplitude, making the ansatz true by definition at one loop and highly non-trivial at higher loops. The formula (\ref{BDSansatz}) was conjectured after direct calculations of the four-point amplitude to two loops \cite{Anastasiou:2003kj} and three loops \cite{Bern:2005iz}. It was found to be consistent with the five-point amplitude at two loops \cite{Cachazo:2006tj,Bern:2006vw} and three loops \cite{Spradlin:2008uu}.

Now let us consider Wilson loops on the polygon contour (\ref{Wilsonloop}).
A lot is known about the structure of such Wilson loops. In particular, due to the cusps on the contour at the points $x_i$ the Wilson loop is ultraviolet divergent. The divergences of such Wilson loops are intimately related to the infrared divergences of scattering amplitudes \cite{Ivanov:1985np,Korchemsky:1992xv,Korchemskaya:1996je}. Indeed the leading ultraviolet divergence is again the cusp anomalous dimension and one can write an equation very similar to that for the loop corrections to the MHV amplitude,
\begin{equation}
{\rm log}\, W_n = \sum_{l=1}^{\infty} a^l \biggl[ \frac{\Gamma_{\rm cusp}^{(l)}}{(l\epsilon)^2} + \frac{\Gamma^{(l)}}{l \epsilon} \biggr] \sum_i (-\mu_{UV}^2 x_{i,i+2}^2)^{l\epsilon} + F^{\rm WL}_n(x_1,\ldots,x_n;a) + O(\epsilon).
\label{logWn}
\end{equation}

The objects of most interest to us here are the two functions $F_n^{\rm MHV}$ from (\ref{logMn}) and $F_n^{\rm WL}$ from (\ref{logWn}) describing the finite parts of the planar MHV amplitude and Wilson loop respectively. In fact there is by now a lot of evidence that in the planar theory, the two functions are identical up to an additive constant,
\begin{equation}
F_n^{\rm MHV}(p_1,\ldots,p_n;a) = F_n^{\rm WL}(x_1,\ldots,x_n;a) + d_n(a)
\label{MHV/WLduality}
\end{equation}
once one changes variables from $p_i$ to $x_i$ as in (\ref{dualcoords}).

The relation between amplitudes and Wilson loops first appeared at strong coupling \cite{Alday:2007hr}. In this regime the identification is a consequence of a particular T-duality transformation of the string sigma model which maps the AdS background into a dual AdS space \cite{Kallosh:1998ji}. The calculation of the scattering amplitude is then reduced to a minimal surface calculation in AdS space with the boundary of the surface being the light-like polygon on the boundary of AdS. The infinite coupling set up does not distinguish between different helicity configurations; the distinction should become apparent when subleading corrections are taken into account.
The identification between amplitudes and Wilson loops is also true in the perturbative regime, however here it is important that the amplitudes are MHV. This is why we have referred only to the MHV amplitudes in the description of the duality. The fact that the Wilson loop is dual to the MHV amplitudes is in some sense natural because both are described by a single scalar function of the kinematic variables, making the identification possible. Non-MHV amplitudes have a richer structure which is so far not incorporated into the duality.
The fact that the duality between planar MHV amplitudes and Wilson loops holds at both strong and weak coupling suggests that it should hold non-perturbatively. The explicit matching of the two quantities was observed at four points and one loop \cite{Drummond:2007aua} and generalised to $n$ points in \cite{Brandhuber:2007yx}. Two loop calculations then followed \cite{Drummond:2007cf,Drummond:2007au,Drummond:2007bm,Drummond:2008aq,Bern:2008ap}. In each case the duality relation (\ref{MHV/WLduality}) was indeed verified.

A point to be stressed here is that dual conformal symmetry finds a natural home within the duality between amplitudes and Wilson loops. It is simply the ordinary conformal symmetry of the Wilson loop defined in the dual space. Moreover, since this symmetry is a Lagrangian symmetry from the point of view of the Wilson loop, its consequences can be derived in the form of Ward identities \cite{Drummond:2007cf,Drummond:2007au}. We gave the form of the generator of special conformal transformations in (\ref{dualK}). Here we keep only the part acting on the $x_i$ as the Wilson loop is only a function of these variables.
\begin{equation}
K_\mu = \sum_i \biggl[ x_{i\mu} x_i \cdot \frac{\partial}{\partial x_i} - \frac{1}{2} x_i^2 \frac{\partial}{\partial x_i^\mu}\biggr].
\end{equation}
The analysis of \cite{Drummond:2007au} shows that the ultraviolet divergences induce an anomalous behaviour for the finite part $F_n^{\rm WL}$ which is entirely captured by the following conformal Ward identity
\begin{equation}
K^\mu F_n^{\rm WL} = \frac{1}{2} \,\, \Gamma_{\rm cusp}(a) \sum_i (2x_i^\mu - x_{i-1}^\mu -x_{i+1}^\mu)\log x_{i-1,i+1}^2.
\label{CWI}
\end{equation}

A very important consequence of the conformal Ward identity is that the finite part of the Wilson loop is fixed up to a function of conformally invariant cross-ratios,
\begin{equation}
u_{ijkl} = \frac{x_{ij}^2 x_{kl}^2}{x_{ik}^2 x_{jl}^2}.
\end{equation}
In the cases of four and five edges, there are no such cross-ratios available due to the light-like separations of the cusp points. This means that the conformal Ward identity (\ref{CWI}) has a unique solution up to an additive constant.
Remarkably, the solution coincides with the Bern-Dixon-Smirnov all-order ansatz for the corresponding scattering amplitudes (here we give the formulas in terms of the $x_i$ variables),
\begin{align}\label{n45}
F_4^{\rm (BDS)} &= \frac{1}{4} \Gamma_{\rm cusp}(a) \log^2 \Bigl(
\frac{x_{13}^2}{x_{24}^2} \Bigr)+ \text{ const },\\
F_5^{\rm (BDS)} &= -\frac{1}{8} \Gamma_{\rm cusp}(a) \sum_{i=1}^5 \log \Bigl(
\frac{x_{i,i+2}^2}{x_{i,i+3}^2} \Bigr) \log \Bigl(
\frac{x_{i+1,i+3}^2}{x_{i+2,i+4}^2} \Bigr) + \text{ const }. \label{n45'}
\end{align}
Thus, taking the Ward identity over from the Wilson loops to the MHV amplitudes, we see that the agreement of the amplitudes with the BDS ansatz for four and five points can be explained by dual conformal symmetry.

In fact the BDS ansatz provides a particular solution to the conformal Ward identity for any number of points. From six points onwards however the functional form is not uniquely fixed as there are conformal cross-ratios available. 
Thus the general solution of the Ward identity contains an arbitrary function of the cross-ratios (which can also depend on the coupling),
\be
F_n^{\rm (WL)} = F_n^{\rm (BDS)} + f_n(u_i;a).
\label{solutionWI}
\ee
At six points, for example there are three cross-ratios,
\begin{equation}
u_1 = \frac{x_{13}^2 x_{46}^2}{x_{14}^2 x_{36}^2}, \qquad u_2 = \frac{x_{24}^2 x_{51}^2}{x_{25}^2 x_{41}^2}, \qquad u_3 = \frac{x_{35}^2 x_{62}^2}{x_{36}^2 x_{41}^2}.
\end{equation}
The solution to the Ward identity is therefore
\begin{equation}
F_6^{\rm (WL)} = F_6^{\rm (BDS)} + f_6(u_1,u_2,u_3;a)\ .
\end{equation}
Here, upon the identification $p_i=x_i-x_{i+1}$,
\begin{align}\label{BDS6point}
  F_{6} ^{\rm (BDS)}   =  \frac{1}{4} \Gamma_{\rm
 cusp}(a)\sum_{i=1}^{6} &\bigg[
    - \log \Bigl(
\frac{x_{i,i+2}^2}{x_{i,i+3}^2} \Bigr)\log\Bigl(
\frac{x_{i+1,i+3}^2}{x_{i,i+3}^2} \Bigr)
\nonumber\\
 &  +\frac{1}{4} \log^2 \Bigl( \frac{x_{i,i+3}^2}{x_{i+1,i+4}^2}
\Bigr)   -\frac{1}{2} {\rm{Li}}_{2}\Bigl(1-
 \frac{x_{i,i+2}^2  x_{i+3,i+5}^2}{x_{i,i+3}^2 x_{i+2,i+5}^2}
\Bigr) \bigg] + \text{ const }\,,
\end{align}
while $f_6(u_1,u_2,u_3;a)$ is some function of the three
cross-ratios and the coupling. The function $f_6$ is not fixed by the Ward identity and has to be determined by explicit calculation of the Wilson loop. The calculation of \cite{Brandhuber:2007yx} shows that at one loop $f_6$ is constant (recall that at one loop the BDS ansatz is true by definition and the Wilson loop and MHV amplitude are known to agree for an arbitrary number of points). At two loops, direct calculation shows that $f_6$ is a non-trivial function \cite{Drummond:2007bm,Drummond:2008aq}. Moreover the calculation \cite{Bern:2008ap} of the six-particle MHV amplitude shows explicitly that the BDS ansatz breaks down at two loops and the same function appears there,
\begin{equation}
F_6^{\rm MHV} = F_6^{\rm WL} + const,\qquad F_6^{\rm MHV} \neq F_6^{\rm BDS}.
\end{equation}
The agreement between the two functions $F_6^{\rm MHV}$ and $F_6^{\rm WL}$ was verified numerically to within the available accuracy. Subsequently the integrals appearing in the calculation of the finite part of the of the hexagonal Wilson loop have been evaluated analytically in terms of multiple polylogarithms \cite{DelDuca:2010zg}.

Further calculations of polygonal Wilson loops have been performed. The two-loop diagrams appearing for an arbitrary number of points have been described in \cite{Anastasiou:2009kna} where numerical evaluations of the seven-sided and eight-sided light-like Wilson loops were made. These functions have not yet been compared with the corresponding MHV amplitude calculations \cite{Vergu:2009zm,Vergu:2009tu} due the difficulty of numerically evaluating the relevant integrals. However given the above evidence it seems very likely that the agreement between MHV amplitudes and light-like polygonal Wilson loops will continue to an arbitrary number of points, to all orders in the coupling.

While the agreement between Wilson loops is fascinating it is clearly not the end of the story. In particular the duality applies only to the MHV amplitudes. In the strict strong coupling limit this does not matter since all amplitudes are dominated by the minimal surface in AdS, independently of the helicity configuration \cite{Alday:2007hr}. At weak coupling that is certainly not the case and non-MHV amplitudes reveal a much richer structure than their MHV counterparts which is so far not captured by any dual object like a Wilson loop. 

Despite the absence of such a dual model, one may still ask what happens to dual conformal symmetry for non-MHV amplitudes. 
Based on analysis of the one-loop NMHV amplitudes it was conjectured in \cite{Drummond:2008vq} that the dual conformal anomaly is universal, i.e. is independent of the MHV degree. This is very natural because we have seen that the anomaly arises due to the infrared divergences (or ultraviolet divergences of the Wilson loop) and these are known to be independent of the helicities of the incoming particles. This means that one can write the all-order superamplitude as a product of the MHV superamplitude and an infrared finite ratio function,
\begin{equation}
\mathcal{A}_n = \mathcal{A}_n^{\rm MHV} R_n.
\label{ratiofunc}
\end{equation}
The conjecture of universality of the anomaly states that, setting the regulator to zero, $R_n$ is dual conformally invariant,
\begin{equation}
K^\mu R_n = 0.
\label{nonMHVinvariance}
\end{equation}
In \cite{Drummond:2008bq} it was argued that this conjecture holds for NMHV amplitudes at one loop, based on explicit calculations up to nine points using supersymmetric generalised unitarity. Subsequently \cite{Brandhuber:2009xz,Elvang:2009ya,Brandhuber:2009kh} it has been argued to hold for all one-loop amplitudes by analysing the dual conformal anomaly arising from infrared divergent two-particle cuts.
 
Note that the conjecture (\ref{nonMHVinvariance}) makes reference only to the dual conformal generator $K$ and not to the full set of dual superconformal transformations. The reason is that some of these transformations overlap with the broken part of the original superconformal symmetry. In particular the generator $\bar{Q}$ does not annihilate the ratio function $R_n$. This fact is related to the breaking of the original superconformal invariance by loop corrections since $\bar{Q}$ is really the same symmetry as $\bar{s}$. Indeed, even at tree-level $\bar{s}$ is subtly broken by contact term contributions \cite{Bargheer:2009qu,Korchemsky:2009hm,Sever:2009aa}. At one loop unitarity relates the discontinuity of the amplitude in a particular channel to the product of two tree-level amplitudes integrated over the allowed phase space of the exchanged particles (as illustrated in Fig. \ref{2cut}). The subtle non-invariance of the trees therefore translates into non-invariance of the discontinuity and therefore of the loop amplitude itself \cite{Korchemsky:2009hm,Sever:2009aa,Beisert:2010gn}. This breaking of the $\bar{s}=\bar{Q}$ symmetry is in addition to that induced by the IR divergences as can be seen by considering a unitarity cut with more than two particles on each side. Such a cut is IR finite but not $\bar{Q}$ invariant as discussed in \cite{Korchemsky:2009hm,Beisert:2010gn}.

\begin{figure}
\psfrag{hat1}[cc][cc]{$\hat{1}$}
\psfrag{hat4}[cc][cc]{$\hat{n}$}
\psfrag{2}[cc][cc]{$i-1$}
\psfrag{3}[cc][cc]{$i$}
\psfrag{MHVb}[cc][cc]{}
\psfrag{MHV}[cc][cc]{}
 \centerline{{\epsfysize4cm
\epsfbox{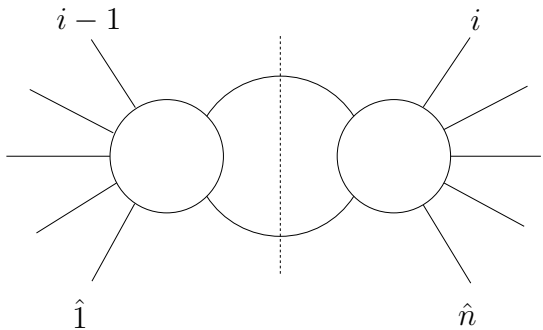}}}  \caption[]{\small A unitarity cut of a one-loop amplitude is given by the product of two tree-level amplitudes integrated over the phase space of the two exchanged on-shell particles.}
  \label{2cut}
\end{figure}

\section{Summary}

We have seen how the use of general considerations about the analytic structure of scattering amplitudes has led to a lot of insights into the nature of the S-matrix of gauge theories. The BCFW recursion relations allow for simple explicit expressions for tree-level scattering amplitudes. With the addition of maximal supersymmetry they become an extremely powerful tool. The expressions obtained contain non-local poles despite the fact that the underlying gauge theory is local. The non-local poles are spurious, cancelling between the different terms in the full amplitude. Their presence is connected to the fact that there are non-local symmetries at work. Each term in the BCFW expansion is an invariant of two copies of superconformal symmetry, the original symmetry of the Lagrangian and a new dual superconformal symmetry. Together these symmetries generate an infinite-dimensional Yangian symmetry. On the S-matrix the original superconformal symmetry acts locally but the extra charges generated by the dual superconformal symmetry act non-locally. Thus to express the amplitude in a way which reveals the symmetry inevitably means that locality is not manifest. As discussed in \cite{ArkaniHamed:2008gz} there are good reasons for wanting a formulation of the S-matrix which does not refer to spacetime locality in an intrinsic way, in particular if one wants to include gravitational physics in the picture. These ideas suggest that the nicest formulation of the S-matrix of quantum field theories in general should not be intrinsically local and that the observation of the additional symmetries in the $\mathcal{N}=4$ case is perhaps the most concrete manifestation of a general principle.

Beyond tree level the dual conformal symmetry continues to provide strong constraints on the form of planar scattering amplitudes. Indeed the MHV amplitudes are equivalently described by light-like Wilson loops in the dual space whose conformal symmetry is the dual conformal symmetry of the amplitudes. Since the Wilson loops obey a conformal Ward identity, they and the corresponding MHV amplitudes are fixed up to a function of (dual) conformal invariants. This is sufficient to fix the four-point and five-point Wilson loops and amplitudes to all orders.

There are many open directions to pursue. A general picture that has emerged over recent years is that planar $\mathcal{N}=4$ super Yang-Mills theory is governed by some integrable structure. Indeed there is strong evidence that the spectrum of anomalous dimensions in the theory can be obtained from a nested Bethe ansatz \cite{Beisert:2005fw} or Y-system \cite{Gromov:2009tv}. These systems of equations arise in physical models from some underlying quantum group structure. It is known that the classical sigma model does indeed exhibit an infinite-dimensional symmetry algebra \cite{Bena:2003wd}, arising from the existence of a one-parameter family of flat connections. In the study of scattering amplitudes (or Wilson loops) at strong coupling this structure has been exploited to derive systems of equations similar to those arising in the spectral problem \cite{Alday:2010vh}. While the existence of extra symmetries has been observed at weak coupling, as we have discussed in these notes, it is not yet known how to tie it in concretely with the above ideas. Thus one of the most central questions is how to exploit the integrability of theory to tell us more about the loop corrections. For example can we fix the remainder function $f_n$ of (\ref{solutionWI}) using the symmetries? Perhaps one should talk directly about the ratio function $R_n$ of (\ref{ratiofunc}). Both of these quantities are infrared finite and are presumably controlled by the integrable structure lying behind the theory. 
At present there is not a complete understanding of the nature of the breaking of the original conformal symmetry by the amplitudes beyond tree-level. This has prevented it being used as a predictive tool for understanding the form of the remainder function $f_n$ or the ratio function $R_n$. A related question is whether there is a generalisation of the amplitude/Wilson loop duality beyond the MHV amplitudes. Such a model should provide some understanding of the breaking of the $\bar{Q}$ supersymmetry observed in the amplitudes. For recent developments on these issues see \cite{Alday:2010zy,Eden:2010zz,ArkaniHamed:2010kv,Drummond:2010mb,Mason:2010yk,Eden:2010ce,CaronHuot:2010ek}.

\section*{Acknowledgements}
I would like to thank Livia Ferro, Johannes Henn, Gregory Korchemsky, Jan Plefka, Vladimir Smirnov and Emery Sokatchev for collaboration on the topics presented in these notes. I also thank Johannes Henn for comments on the manuscript and the organisers of the CERN winter school for the opportunity to lecture at such an enjoyable school.

\appendix

\setcounter{section}{0} \setcounter{equation}{0}
\renewcommand{\theequation}{\Alph{section}.\arabic{equation}}

\section{Conventions}
We use the following conventions for the spinor contractions. A vector can be exchanged for a bispinor making use of the spin matrices $\sigma_\mu$,
\be
x^{\a \adt} = (\sigma_\mu)^{\a \adt} x^\mu, \qquad x^2 = x^\mu x_\mu = 2 x^{\a \adt} x_{\a \adt}.
\ee
Derivatives are defined by
\be
\partial_{\a \adt} x^{\b \bdt} = \delta_\a^\b \delta_\adt^{\bdt}, \qquad \partial_\a \lambda^\b = \delta_\a^\b \text{ etc. }
\ee
The spinor scalar products are defined as follows,
\be
\l< a b \r> = a^\a b_\a = a^\a b^\b \epsilon_{\b \a}, \qquad [a b] = a_\adt b^\adt = a_\adt b_\bdt \epsilon^{\adt \bdt}
\ee
Longer spinor contractions are also used
\be
\l< a | P | b] = a^\a P_{\a \adt} b^{\adt}, \qquad \l< a| x y |b\r> = a^\a x_{\a \adt} y^{\adt \b} b_{\b}, 
\ee
etc. 

We now give the commutation relations for the superconformal algebra.
We begin by listing the commutation relations of the algebra $u(2,2|4)$. The Lorentz generators
$\mathbb{M}_{\a \b}$, $\overline{\mathbb{M}}_{\adt \bdt}$ and the $su(4)$ generators
$\mathbb{R}^{A}{}_{B}$ act canonically on the remaining generators carrying Lorentz or $su(4)$
indices. The dilatation $\mathbb{D}$ and hypercharge $\mathbb{B}$ act via
\be
[\mathbb{D},\mathbb{J}] = {\rm dim}(\mathbb{J})\, \mathbb{J}, \qquad [\mathbb{B},\mathbb{J}] = {\rm
hyp}(\mathbb{J})\, \mathbb{J}.
\ee
The non-zero dimensions and hypercharges of the various generators are
\begin{align} \notag
& {\rm dim}(\mathbb{P})=1, \qqquad {\rm dim}(\mathbb{Q}) = {\rm dim}(\overline{\mathbb{Q}}) =
\tfrac{1}{2},\qquad {\rm dim}(\mathbb{S}) = {\rm dim}(\overline{\mathbb{S}}) = -\tfrac{1}{2}
\\
&{\rm dim}(\mathbb{K})=-1,\qquad {\rm hyp}(\mathbb{Q}) = {\rm hyp}(\overline{\mathbb{S}}) =
\tfrac{1}{2}, \qquad~ {\rm hyp}(\overline{\mathbb{Q}}) = {\rm hyp}(\mathbb{S}) = - \tfrac{1}{2}.
\end{align}
The remaining non-trivial commutation relations are,
\begin{align} \notag
& \{\mathbb{Q}_{\a A},\overline{\mathbb{Q}}_{\adt}^B\}  =  \delta_A^B \mathbb{P}_{\a \adt},
   \qquad \{\mathbb{S}_{\a}^A,\overline{\mathbb{S}}_{\adt B} \} = \delta_B^A \mathbb{K}_{\a \adt},
\\ \notag
& {}[\mathbb{P}_{\a \adt},\mathbb{S}^{\b A}] = \delta_{\a}^{\b} \overline{\mathbb{Q}}_{\adt}^A,
 \qqquad [\mathbb{K}_{\a \adt},\mathbb{Q}^{\b}_{A}] = \delta_{\a}^{\b}
   \overline{\mathbb{S}}_{\adt A},
\\ \notag
& {}[\mathbb{P}_{\a \adt},\overline{\mathbb{S}}^{\bdt}_{A}]  =  \delta^{\bdt}_{\adt} \mathbb{Q}_{\a A},
\qqquad [\mathbb{K}_{\a \adt}, \overline{\mathbb{Q}}^{\bdt A}]  =  \delta_{\adt}^{\bdt} \mathbb{S}_{\a}^{A},
\\ \notag
& [\mathbb{K}_{\a \adt},\mathbb{P}^{\b \bdt}] = \delta_\a^\b \delta_\adt^\bdt \mathbb{D} +
\mathbb{M}_{\a}{}^{\b}
 \delta_\adt^\bdt + \overline{\mathbb{M}}_{\adt}{}^{\bdt} \delta_\a^\b,
\\ \notag
& \{\mathbb{Q}^{\a}_{A},\mathbb{S}_\b^B\} =  \mathbb{M}^{\a}{}_{\b}
\delta_A^B + \delta^{\a}_{\b} \mathbb{R}^{B}{}_{A} + \tfrac{1}{2}\delta^{\a}_{\b} \delta_A^B (\mathbb{D}+\mathbb{C}),
\\ 
& \{\overline{\mathbb{Q}}^{\adt A},\overline{\mathbb{S}}_{\bdt B}\} = \overline{\mathbb{M}}^{\adt}{}_{\bdt} \delta_B^A  - \delta^{\adt}_{\bdt} \mathbb{R}^{A}{}_{B} + \tfrac{1}{2} \delta^{\adt}_{\bdt}\delta_B^A
(\mathbb{D}-\mathbb{C}).
\label{comm-rel}
\end{align}
Removing the hypercharge yields $su(2,2|4)$. Setting the central charge to zero gives $pu(2,2|4)$. Doing both gives the superconformal algebra $psu(2,2|4)$.

\bibliography{lectures}
\bibliographystyle{nb}

\end{document}